\documentclass[aps,prd,twocolumn,superscriptaddress,nofootinbib,natbib]{revtex4-2}
\usepackage[T1]{fontenc}
\usepackage{amsmath}
\usepackage{booktabs}
\usepackage{xcolor}
\usepackage{hyperref}
\hypersetup{colorlinks=true, linkcolor=blue, citecolor=blue, urlcolor=blue}

\usepackage{placeins}
\usepackage{comment}
\usepackage{graphicx}
\usepackage{grffile}
\usepackage{tikz}

\newcommand{\safefig}[2][width=\columnwidth]{%
  \IfFileExists{#2}{%
    \includegraphics[#1]{#2}%
  }{%
    \fbox{\parbox{0.8\columnwidth}{\centering\vspace{2cm}%
    {\small\texttt{#2}\\[4pt] Figure not yet generated}%
    \vspace{2cm}}}%
  }%
}

\newcommand{\Msun}{M_\odot}
\newcommand{\Lgamma}{L_\gamma}
\newcommand{\thetaobs}{\theta_{\rm obs}}
\newcommand{\Ye}{Y_e}

\newcommand{\ds}{\mathrm{d}s}

\begin{document}

\title{Gamma-ray Signatures of  $r$-Process Radioactivity from the Collapse of Magnetized White Dwarfs}

\author{Tetyana Pitik}
\email{tetyana.pitik@berkeley.edu}
\affiliation{Department of Physics, University of California Berkeley, Berkeley, California 94720, USA}

\author{Yong-Zhong Qian}
\affiliation{School of Physics and Astronomy, University of Minnesota, Minneapolis, Minnesota 55455, USA}

\author{David Radice}
\affiliation{Department of Astronomy \& Astrophysics, The Pennsylvania State University, University Park PA 16802, USA}
\affiliation{Institute for Gravitation and the Cosmos, The Pennsylvania State University, University Park PA 16802, USA}
\affiliation{Department of Physics, The Pennsylvania State University, University Park PA 16802, USA}

\author{Daniel Kasen}
\affiliation{Department of Physics, University of California Berkeley, Berkeley, California 94720, USA}
\affiliation{Department of Astronomy, University of California Berkeley, Berkeley, California 94720, USA}
\affiliation{Nuclear Science Division, Lawrence Berkeley National Laboratory, Berkeley, CA 94720, USA}

\date{\today}

\begin{abstract}
We predict the gamma-ray line emission from $r$-process nuclei synthesized in the ejecta of the accretion-induced collapse (AIC) of a magnetized, rapidly rotating white dwarf.  Using ejecta from a two-dimensional general-relativistic neutrino-magnetohydrodynamic simulation, further evolved with a radiation-hydrodynamics code coupled to an in-situ nuclear reaction network, we construct angle-dependent gamma-ray spectra in the $0.01$--$10\,\mathrm{MeV}$ band via composition-dependent ray-tracing through the ejecta.  The emission between $\sim$1 and $10\,$d is dominated by $^{132}$I ($t_{1/2} = 2.3\,$h), continuously replenished by the decay of its parent $^{132}$Te ($t_{1/2} = 3.2\,$d), with additional contributions from $^{131}$I, $^{133}$Xe, and $^{132}$Te.  At $t\gtrsim 20$~d, $^{56}$Co (from $^{56}$Ni decay) becomes the primary emitter.  The simultaneous presence of $r$-process and iron-peak gamma-ray lines is distinctive of AIC ejecta and absent in binary neutron star mergers, where iron-peak nuclei are generally not synthesized.  Comparing with the $3\sigma$ continuum sensitivities of planned MeV gamma-ray telescopes (COSI, AMEGO-X, e-ASTROGAM, GRAMS, GammaTPC), we find the brightest $r$-process lines detectable to $\sim 10\,\mathrm{Mpc}$ by GammaTPC and GRAMS, with the signal approaching their sensitivity threshold at $30\,\mathrm{Mpc}$.  The $r$-process spectral features survive time integration over $\sim$30\,d exposures, demonstrating robustness against the long observation times required by gamma-ray detectors.
\end{abstract}

\maketitle



\section{INTRODUCTION}
\label{sec:intro}

White dwarfs (WDs) are the most common stellar remnants in the Galaxy, representing the end state of low- and intermediate-mass stars.  Depending on the progenitor mass and evolution, these compact objects are composed predominantly of carbon-oxygen (CO) or oxygen-neon (ONe).  When a WD in a binary system accretes mass from a companion star and approaches the Chandrasekhar limit, its fate diverges sharply depending on its composition.  CO WDs are generally believed to undergo thermonuclear runaway and explode as Type Ia supernovae, leaving no remnant \citep{2007MNRAS.380..933Y, 2009ApJ...692..324S, 2009ApJ...705..693S, 2013ApJ...776...97M}.  In contrast, more massive ONe WDs can collapse into neutron stars (NSs) as electron-capture reactions on neon and magnesium trigger the loss of pressure support once the Chandrasekhar mass is exceeded \citep{1986PrPNP..17..249N, 2019ApJ...886...22Z}.  This process -- the \emph{accretion-induced collapse} (AIC) of a WD -- was first proposed by \citep{1991ApJ...367L..19N} and has since been studied with increasingly sophisticated numerical simulations \citep{1999ApJ...516..892F, Dessart:2006pe,2007ApJ...669..585D, Abdikamalov:2009aq, 2023MNRAS.525.6359L, Cheong:2024hrd, Batziou:2024ory,Kuroda:2025iyj,Combi:2025yvs,Pitik:2026bjm}. A similar outcome can occur through the merger of two WDs (merger-induced collapse, MIC), where the combined mass exceeds the stability threshold~\citep{1985A&A...150L..21S, 2006MNRAS.368L...1L}.

The AIC channel is of particular interest for several reasons.  First, it provides a pathway to form NSs without a core-collapse supernova, potentially contributing to the population of millisecond pulsars in globular clusters and the formation of low-kick NS binary systems \citep{1984JApA....5..209V, Freire:2013xma,Wang:2020pzc,Kremer:2023utr}.  Second, general-relativistic magnetohydrodynamic (GRMHD) simulations have shown that, when the collapsing WD carries a sufficiently strong magnetic field, the resulting proto-neutron star can launch magnetically driven outflows that eject $\sim \mathcal{O}(10^{-1}\,\Msun)$ of neutron-rich material on dynamical timescales \citep{2007ApJ...669..585D, Cheong:2024hrd,Pitik:2026bjm}.

In a companion paper \citep{Pitik:2026bjm}, we presented the first end-to-end calculation of this process, from the magnetized AIC collapse through $r$-process nucleosynthesis to kilonova-like electromagnetic emission. We found that the ejected neutron-rich outflow ($\langle \Ye \rangle \sim 0.24$) undergoes strong  $r$-process nucleosynthesis, producing heavy elements up to and beyond the third  $r$-process peak ($A \geq 195$), including a substantial lanthanide mass fraction ($X_{\rm lan} \sim 6\%$), and demonstrated striking agreement with the observations of the kilonova AT\,2023vfi associated with the long-duration gamma-ray burst GRB\,230307A \citep{Levan:2024, 2022Natur.612..223R}.  Those results established magnetized AIC as a viable progenitor channel for $r$-process element production and long-duration GRBs accompanied by kilonova signatures.

In this work, we focus on the gamma-ray line emission from the radioactive decay of these freshly synthesized $r$-process nuclei.  Gamma-ray observations offer a unique and direct diagnostic of the nuclear composition of the ejecta, as each radioactive isotope produces a characteristic set of gamma-ray lines with known energies and branching ratios.  Unlike the optical/infrared kilonova signal, which depends on the complex interplay of opacities, thermalization, and radiative transfer, gamma-ray lines can in principle directly reveal the identity and abundance of specific isotopes.  This approach has been explored in the context of binary neutron star (BNS) mergers, both for the early kilonova phase \citep{Hotokezaka:2015cma, Li:2019gamma, Korobkin:2019uxw, Chen:2021tob, Magistrelli:2025xja} and for ancient Galactic remnants \citep{Terada:2022hut, Gross:2025yff}, but has not yet been applied to AIC ejecta.

The detectability of these gamma-ray signatures depends on several factors: the total mass and composition of the ejecta, the time-dependent opacity to gamma-rays (which determines the escape fraction), the viewing angle (since AIC ejecta are highly aspherical), and the distance to the source.  Planned and upcoming missions such as the Compton Spectrometer and Imager \citep[COSI;][]{Tomsick:2023aue}, the All-sky Medium Energy Gamma-ray Observatory \citep[AMEGO-X;][]{Caputo:2022xpx}, e-ASTROGAM \citep{e-ASTROGAM:2017pxr}, GammaTPC \citep{Shutt:2025xvc}, and Gamma-Ray and AntiMatter Survey \citep[GRAMS;][]{GRAMS:2025ljc}, provide sensitivity in the $\sim 0.01$--$10\,\mathrm{MeV}$ energy band that is relevant for nuclear gamma-ray lines.

The paper is organized as follows.  In Section~\ref{sec:simulation}, we briefly summarize the simulation setup and ejecta properties from our companion paper.  In Section~\ref{sec:methods}, we describe the gamma-ray emission calculation, including the nuclear decay data, the opacity model, and the angle-dependent ray-tracing procedure.  Section~\ref{sec:results} presents our results: the time-dependent gamma-ray luminosity, the dominant contributing isotopes, the angle-resolved spectra, and the comparison with instrument sensitivities. We discuss the implications in Section~\ref{sec:discussion} and summarize our conclusions in Section~\ref{sec:conclusions}.

\section{SIMULATION OVERVIEW}
\label{sec:simulation}

Here we briefly summarize the simulation framework used to generate the ejecta profiles employed in this work. We refer the reader to \citep{Pitik:2026bjm} for a more detailed description of the simulation setup and results. This simulation employs equal poloidal and toroidal magnetic field strengths $B_{\rm pol} = B_{\rm tor} = 10^{12}\,\mathrm{G}$.  In \citep{Cheong:2024hrd}, we explored four values of $B_{\rm pol} = \{10^{9}, 10^{10}, 10^{11}, 10^{12}\}\,$G in 2D axisymmetric simulations, and only the model with the strongest field led to the ejection of a significant mass of neutron-rich ejecta.  This was the motivation to use this model in \citep{Pitik:2026bjm} to study the nucleosynthetic outcome and kilonova signal.
The large initial field adopted in 2D serves as a proxy for the magnetic field amplification by the magnetorotational instability (MRI) and turbulent dynamo, which operate in the post-bounce accretion disk but cannot be captured self-consistently in axisymmetric simulations due to the anti-dynamo theorem.  The 3D GRMHD simulations of \citep{Combi:2025yvs} confirm that even starting from a much weaker seed field confined below the WD surface, an MRI-driven $\alpha\Omega$ dynamo in the accretion disk can rapidly amplify the magnetic energy by many orders of magnitude within ${\sim} 50\,$ms, launching a jet comparable to those obtained in our 2D setup.  This confirms that the strong-field initial conditions in 2D are a reasonable representation of the physical outcome, rather than an extreme assumption.  Whether weaker initial fields in 3D --- where the dynamo can operate self-consistently --- could lead to similarly neutron-rich ejecta remains an open question that we plan to investigate in future 3D simulations.

The collapse of the magnetized WD is simulated in two-dimensional axisymmetry using the general-relativistic neutrino-MHD code \texttt{Gmunu} \citep{2020CQGra..37n5015C, 2021MNRAS.508.2279C}.  The initial model is a rigidly rotating WD with gravitational mass $1.5\,\Msun$, and axis ratio $a_r = 0.75$ ($\Omega \sim 5\,\mathrm{Hz}$). The simulation employs the LS220 equation of state \citep{1991NuPhA.535..331L} and an energy-integrated two-moment neutrino transport scheme.  The computational domain extends to $3 \times 10^5\,\mathrm{km}$ with 14 levels of adaptive mesh refinement and a minimum vertical spacing of $\Delta z \simeq 286\,\mathrm{m}$ near the stellar centre.  Both hemispheres are evolved independently without imposing equatorial symmetry.

Following core bounce, the strong magnetic field drives magnetically dominated outflows along the polar axis, while simultaneously accelerating uncollimated ejecta at mid-latitudes.  The total ejected mass is $M_{\rm ej} \approx 0.18\,\Msun$ with kinetic energy $E_{\rm ej} \simeq 8 \times 10^{50}\,\mathrm{erg}$.  The ejecta distribution is highly anisotropic: the bulk of the neutron-rich material ($\Ye \lesssim 0.25$), which constitutes ${\sim} 0.14\,\Msun$, is concentrated at mid-latitudes ($\theta \sim 20^\circ$--$85^\circ$ and $95^\circ$--$160^\circ$) with typical velocities $v_\infty \lesssim 0.2\,c$ and low specific entropies ($s \lesssim 20\,k_B/\mathrm{baryon}$), while higher-$\Ye$ material resides near the equatorial plane and in the fast polar outflow.  The mass-weighted averages are $\langle \Ye \rangle \sim 0.24$, $\langle v_\infty \rangle \sim 0.1\,c$, and $\langle s \rangle \sim 9\,k_B/\mathrm{baryon}$.  The condition of low $\Ye$ is particularly favourable for $r$-process nucleosynthesis.

The physics underlying this $\Ye$ distribution is directly tied to the competition between magnetically driven ejection and neutrino irradiation.  During the collapse, electron captures on protons drive the forming accretion disk to very low electron fractions ($\Ye \sim 0.1$).  In unmagnetized or weakly magnetized models, the material remains bound near the proto-neutron star long enough for sustained electron-neutrino irradiation to convert neutrons back to protons, raising $\Ye$ toward ${\sim}\,0.5$ by the time weak interactions freeze out, (see, e.g,~\citep{2009MNRAS.396.1659M}).
In the strongly magnetized model considered here, the rapid magnetic field amplification in the polar funnel and at mid-latitudes launches material on dynamical timescales, before neutrino irradiation can significantly re-leptonize the outflow.  This preserves the low $\Ye$ produced during collapse, particularly at mid-latitudes.  The higher-$\Ye$ material near the equator corresponds to the slower, denser accretion torus outflow that remains exposed to the neutrino field for longer, while the fast polar jet also has increasing $\Ye$ with time because it originates from the immediate vicinity of the proto-neutron star surface where the neutrino flux is most intense (see the discussion and Figs.~2, 3 and 4 of \citep{Pitik:2026bjm} for the detailed $\Ye$--velocity--angle distributions).

At ${\sim} 1\,\mathrm{s}$ post-bounce, ejecta profiles are extracted along 35 polar directions ($\theta = 5^\circ$ to $175^\circ$ in $5^\circ$ increments) and evolved with the one-dimensional Lagrangian radiation-hydrodynamics code \texttt{kNECnn} \citep{Morozova:2015bla, Wu:2021ibi, Magistrelli:2025xja}, which couples the implicit flux-limited diffusion scheme of \texttt{SNEC} to the nuclear reaction network \texttt{SkyNet} \citep{Lippuner:2017tyn}.  Each Lagrangian mass shell independently evolves its own \texttt{SkyNet} instance, tracking 7836 isotopes and ${\sim}140\,000$ nuclear reactions with rates from the JINA REACLIB database \citep{2010ApJS..189..240C} and nuclear masses from the FRDM model \citep{Moller:2015fba},  where experimental masses are not available.  The nucleosynthesis proceeds through qualitatively different channels depending on latitude: the neutron-rich mid-latitude material undergoes strong $r$-process nucleosynthesis, producing heavy elements up to and beyond the third $r$-process peak ($A \gtrsim 195$), while the higher-$\Ye$ equatorial material synthesizes lighter species, including iron-group and first-peak elements.  The resulting mass-weighted abundance pattern shows remarkable agreement with the solar $r$-process residuals across the second ($A \sim 130$) and third ($A \sim 195$) peaks \citep{Pitik:2026bjm}.  The four most abundant elements by mass at $t \simeq 30\,$d are Xe ($Z = 54$), He ($Z = 2$), Pt ($Z = 78$), and Te ($Z = 52$), and the lanthanide mass fraction is $X_{\rm lan} \approx 6\%$.  The large tellurium yield, characteristic of the second $r$-process peak, is directly responsible for the dominant gamma-ray signatures identified in this work.

\section{GAMMA-RAY EMISSION}
\label{sec:methods}

We construct the emergent gamma-ray spectrum as a function of observer time $t$ and polar viewing angle $\thetaobs$ by post-processing the \texttt{SkyNet} nucleosynthesis outputs on a two-dimensional spherical $(r, \theta)$ grid.  The calculation consists of three ingredients: 1) a source term determined by the radioactive decay rates and nuclear gamma-ray line data; 2) an energy-dependent, composition-dependent opacity that governs the escape probability along each line of sight; and 3) the Doppler shift arising from the homologous expansion of the ejecta.  The geometry of the calculation is illustrated in Fig.~\ref{fig:geometry}.\\

\begin{figure}
  \centering
  \begin{tikzpicture}[scale=0.8]

    \fill[blue!12, draw=blue!40, thick, rounded corners=6pt]
        (0.15,0.5) -- (0.2,3.5) -- (1.5,4.3) -- (3.0,3.8)
        -- (4.2,2.2) -- (4.5,1.0) -- (3.0,0.0)
        -- (4.5,-1.0) -- (4.2,-2.2) -- (3.0,-3.8)
        -- (1.5,-4.3) -- (0.2,-3.5) -- (0.15,-0.5) -- cycle;

    \foreach \r in {1.0, 2.0, 3.0, 4.0} {
        \draw[black!19, thin] (0,-\r) arc (-90:90:\r);
    }
    \foreach \ang in {10,20,30,40,50,60,70,80,100,110,120,130,140,150,160,170} {
        \draw[black!19, thin] (0,0) -- ({5.0*sin(\ang)},{5.0*cos(\ang)});
    }

    \node[blue!60, font=\small] at (2.2,2.6) {ejecta};

    \draw[->, thick, black!50] (0,-5.5) -- (0,5.8) node[above] {$z$ (pole)};
    \draw[->, thick, black!50] (0,0) -- (5.2,0) node[right] {$r$ (equator)};

    \def\thobs{25}  
    \draw[->, very thick, red!70!black]
        (0,0) -- ({5.5*sin(\thobs)},{5.5*cos(\thobs)});
    \draw[red!70!black, thick, ->] (0,2.0) arc (90:{90-\thobs}:2.0);
    \node[red!70!black, font=\small] at (0.7,2.4) {$\thetaobs$};
    \node[red!70!black, font=\small] at ({4.5*sin(\thobs)+0.5},{4.5*cos(\thobs)+0.3}) {$\hat{q}$};

    \def\cellang{135}  
    \def\cellr{2.5}    
    \pgfmathsetmacro{\cx}{\cellr*sin(\cellang)}
    \pgfmathsetmacro{\cy}{\cellr*cos(\cellang)}

    \pgfmathsetmacro{\vx}{0.9*sin(\cellang)}
    \pgfmathsetmacro{\vy}{0.9*cos(\cellang)}
    \draw[->, thick, orange!70!black, dashed]
        (\cx,\cy) -- ({\cx+\vx},{\cy+\vy});
    \node[orange!70!black, font=\footnotesize] at ({\cx+\vx+0.3},{\cy+\vy-0.15}) {$\mathbf{v}_{i}$};

    \pgfmathsetmacro{\dx}{sin(\thobs)}
    \pgfmathsetmacro{\dy}{cos(\thobs)}
    \draw[very thick, black]
        (\cx,\cy) -- ({\cx+4.8*\dx},{\cy+4.8*\dy});
    \draw[->, thin, black]
        ({\cx+4.8*\dx},{\cy+4.8*\dy}) -- ({\cx+6*\dx},{\cy+6*\dy});

    \pgfmathsetmacro{\smx}{\cx+1.8*\dx}
    \pgfmathsetmacro{\smy}{\cy+2.0*\dy}
    \node[black, font=\footnotesize, above] at (\smx,\smy) {$s$};

    \pgfmathsetmacro{\arcstart}{90 - \cellang}
    \pgfmathsetmacro{\arcend}{90 - \thobs}
    \draw[black!60, thick, ->]
        ({\cx+0.7*sin(\cellang)},{\cy+0.7*cos(\cellang)})
        arc ({\arcstart}:{\arcend}:0.7);
    \node[black!60, font=\footnotesize] at ({\cx+1.0},{\cy+0.2}) {$\alpha$};

    \fill[orange!80!red] (\cx,\cy) circle (0.12);
    \node[orange!80!red, font=\small, left] at ({\cx-0.02},\cy) { $(r_{i},\theta_{i})$};

    \pgfmathsetmacro{\ex}{\cx+7.2*\dx}
    \pgfmathsetmacro{\ey}{\cy+7.2*\dy}
    \draw[black!70, line width=0.7pt, fill=white]
        ({\ex-0.4},\ey) .. controls ({\ex-0.15},{\ey+0.45}) and ({\ex+0.15},{\ey+0.45}) .. ({\ex+0.4},\ey)
        .. controls ({\ex+0.15},{\ey-0.45}) and ({\ex-0.15},{\ey-0.45}) .. cycle;
    \fill[black!80] (\ex,\ey) circle (0.15);
    \node[font=\small, black!60, below] at (\ex,{\ey-0.5}) {observer};

    \fill[black!60] (0,0) circle (0.15);
    \node[black!60, font=\footnotesize, below left] at (-0.1,-0.2) {remnant};

  \end{tikzpicture}
  \caption{Geometry of the gamma-ray escape calculation.  The ejecta (blue shading) occupy a 2D spherical $(r_{i}, \theta_{i})$ grid covering the full polar range $\theta = 0^\circ$ (pole) to $180^\circ$; only the $r > 0$ meridional half-plane is shown (the configuration is axisymmetric about the $z$-axis).  An emitting cell at position $(r_{i}, \theta_{i})$ radiates gamma-rays toward the observer at polar angle $\thetaobs$.  The bold segment shows the path through the ejecta, along which the element column densities $\Sigma_Z$ (Eq.~\ref{eq:column}) and optical depth $\tau(E)$ (Eq.~\ref{eq:tau_E}) are accumulated; the thin arrow shows the free path outside.  The angle $\alpha$ between the cell velocity $\mathbf{v}_{i} = \mathbf{r}_{i}/t$ and the observer direction $\hat{q}$ determines the Doppler shift (Eq.~\ref{eq:doppler}).}
  \label{fig:geometry}
\end{figure}

Each radioactive isotope $(A,Z)$ in the ejecta decays with mean lifetime $t_{A,Z} = t_{1/2}/\ln 2$ and emits a characteristic set of gamma-ray lines at rest-frame energies $E_{\gamma,k}$ with absolute intensities (photons per decay) $I_{\gamma,k}$.  We take the half-lives and the line data from the Evaluated Nuclear Data File ENDF/B-VIII.0 \citep{2018NDS...148....1B}, which provides $\gamma$-ray emission data for $\sim$1400 radioactive species.

The ejecta are discretized on a two-dimensional spherical grid with cells indexed by $i$, each centred at $(r_i, \theta_i)$.  The angular grid is defined by the 35 \texttt{kNECnn} polar directions ($\theta = 5^\circ$ to $175^\circ$ in $5^\circ$ steps), while a common radial grid of 200 uniform bins extends from the origin to $r_{\rm max} = v_{\rm max}\,t_{\rm ref}$, where $v_{\rm max}$ is determined from the maximum shell velocity across all angular directions at $t_{\rm ref}$.   Each one-dimensional Lagrangian mass shell from a given \texttt{kNECnn} angular direction is distributed across the radial bins proportionally to the volume overlap, conserving mass.  Because each \texttt{kNECnn} direction represents a full-sphere ($4\pi$\,sr) simulation, the shell mass is scaled by the solid-angle fraction $\Delta\Omega/(4\pi) = \sin\theta\,\Delta\theta/2$, where $\Delta\Omega = 2\pi\sin\theta\,\Delta\theta$ is the solid angle of the angular bin (already integrated over the full $2\pi$ in azimuth).  The resulting cell mass $\Delta m_i$ therefore represents the total mass of the azimuthal ring at $(r_i,\theta_i)$.
Since the ejecta expand homologously ($\mathbf{v} = \mathbf{r}/t$) from $t\gtrsim 30$~s after the collapse of the star, shell velocities are constant and the mapping is time-independent.  The composition in each cell is the mass-weighted average over all contributing shells:
\begin{equation}
  Y_{A,Z}^{(i)}(t) = \frac{\sum_j \Delta M_j \, Y_{A,Z}^{(j)}(t)}{\sum_j \Delta M_j},
  \label{eq:Y_cell}
\end{equation}
where $j$ runs over all \texttt{SkyNet} shells overlapping cell $i$ and $\Delta M_j$ is the conservatively remapped mass from shell $j$.  We thus obtain $Y_{A,Z}^{(i)}(t)$ for each isotope in every cell, together with the total conserved cell mass $\Delta m_i = \sum_j \Delta M_j$ and the element mass fractions ${X_Z^{(i)}(t) = \sum_{\{A\}}A\, Y_{A,Z}^{(i)}(t) }$ needed for the opacity calculation.  The instantaneous decay rate per cell of isotope $(A,Z)$ is
\begin{equation}
  \dot{N}^{(i)}_{A,Z} = \frac{\Delta m_{i} \, Y^{(i)}_{A,Z}}{m_u \, t_{A,Z}},
  \label{eq:decay_rate}
\end{equation}
where $m_u$ is the atomic mass unit.  Each decay of isotope $(A,Z)$ releases a total gamma-ray energy 
\begin{equation}
  \varepsilon_{A,Z} = \sum_k I_{\gamma,k}^{(A,Z)} \, E_{\gamma,k}^{(A,Z)},
  \label{eq:eps_decay}
\end{equation}
where the sum runs over all gamma-ray lines $k$ of that isotope, $E_{\gamma,k}^{(A,Z)}$ is the line energy, and $I_{\gamma,k}^{(A,Z)}$ is the absolute intensity (probability of emitting a gamma-ray photon of energy $E_{\gamma,k}^{(A,Z)}$ per decay).  The total intrinsic gamma-ray luminosity of the ejecta is then
\begin{equation}
  \Lgamma^{\rm int}(t) =\sum_i \sum_{A,Z} \dot{N}^{(i)}_{A,Z}\varepsilon_{A,Z},
  \label{eq:Lgamma_int}
\end{equation}
where the sum runs over all radioactive species.

\subsection{Gamma-ray opacity and ray-tracing}
\label{sec:opacity}

At the photon energies of interest ($0.01$--$10\,\mathrm{MeV}$), gamma-rays interact with the ejecta primarily through Compton scattering, photoelectric absorption, and (above the threshold at $1.022\,\mathrm{MeV}$) electron--positron pair production.  We treat all three processes as effective absorption, i.e.\ a photon that undergoes any interaction is removed from the beam.  For photoelectric absorption and pair production this is exact, but for Compton scattering it is an approximation: in reality, the photon is redirected at a lower energy rather than destroyed.  Our treatment therefore neglects the scattered continuum that would result from multiple Compton down-scatterings.    \\
We compute the energy-dependent opacity for each element $Z$ using the \href{https://www.nist.gov/pml/xcom-photon-cross-sections-database}{NIST XCOM} cross-section database, which provides the total effective mass attenuation coefficient $\kappa_Z^{\rm eff}(E_\gamma)$ (in $\mathrm{cm^2\,g^{-1}}$) as a function of photon energy $E_\gamma$.  The XCOM Compton cross sections are computed using Hartree-Fock incoherent scattering functions for isolated neutral atoms \citep{1975JPCRD...4..471H}, which account for all $Z$ electrons of each element.  At gamma-ray energies $E_\gamma \gg E_{\rm bind}$, the incoherent scattering function $S(q,Z) \to Z$, so that all atomic electrons -- whether bound or free -- scatter with approximately the Klein--Nishina cross section.  This is the appropriate regime for the principal gamma-ray lines in our calculation ($E_\gamma \gtrsim 0.2\,\mathrm{MeV}$).

The effective opacity in the cell $i$ of the ejecta is the mass-fraction-weighted sum over all elements:
\begin{equation}
  \kappa^{(i)}_{\rm eff}(E_\gamma) = \sum_Z X^{(i)}_Z \, \kappa_Z^{\rm eff}(E_\gamma),
  \label{eq:kappa_eff}
\end{equation}
where $X^{(i)}_Z$ is the mass fraction of element $Z$ in the cell.  This composition-dependent approach naturally accounts for the higher photoelectric opacity of heavy $r$-process elements ($\sigma_{\rm PE} \propto Z^{4\text{--}5}$), in contrast to the iron-group approximation used in some previous works \citep{Hotokezaka:2015cma, Korobkin:2019uxw}.

\begin{figure*}
  \centering
  \includegraphics[width=0.8\textwidth]{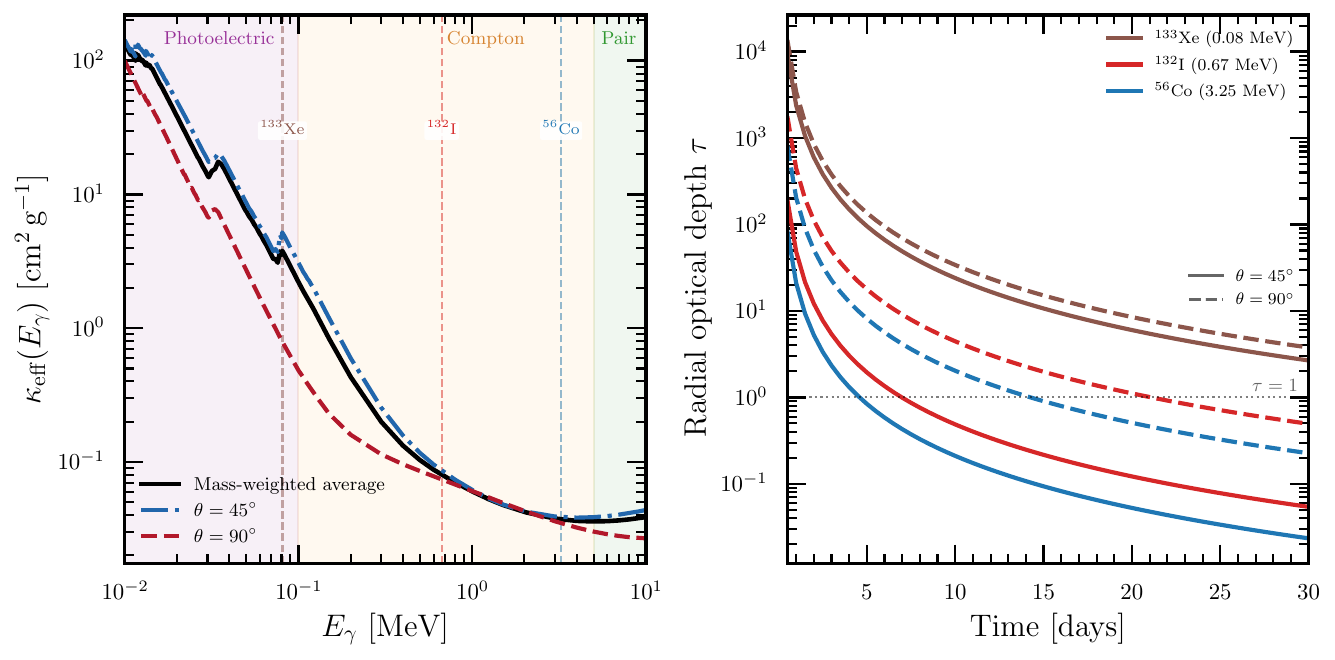}
  \caption{\textit{Left panel (a):} Effective mass attenuation coefficient $\kappa_{\rm eff}(E_\gamma)$ (Eq.~\ref{eq:kappa_eff}) as a function of photon energy, for three representative ejecta compositions: the mass-weighted average over all cells (solid black), $\theta = 45^\circ$ (mid-latitude, $r$-process-rich, dash-dotted blue), and $\theta = 90^\circ$ (equatorial, iron-group-dominated, dashed red).  The three opacity regimes --- photoelectric absorption, Compton scattering, and pair production --- are shaded.  Vertical dashed lines mark the rest-frame energies of $^{133}$Xe ($0.08\,$MeV), $^{132}$I ($0.67\,$MeV), and $^{56}$Co ($3.25\,$MeV).  \textit{Right panel (b):} Radial optical depth $\tau$ (Eq.~\ref{eq:tau_E}) as a function of time for the same three line energies.  Solid and dashed curves show $\tau$ along the $\theta = 45^\circ$ and $90^\circ$ radial directions, respectively.  The horizontal dotted line marks $\tau = 1$.}
  \label{fig:kappa_eff}
\end{figure*}

Figure~\ref{fig:kappa_eff}a shows the effective mass attenuation coefficient $\kappa_{\rm eff}(E_\gamma)$ (Eq.~\ref{eq:kappa_eff}) evaluated for three representative compositions: the mass-weighted average over all ejecta (solid black), and the radially mass-averaged composition at two latitudes --- $\theta = 45^\circ$ (mid-latitude, $r$-process-rich, dash-dotted) and $\theta = 90^\circ$ (equatorial, iron-group-dominated, dashed).  For each curve, the element mass fractions $X_Z$ are averaged over all radial shells at the given latitude, weighted by cell mass, producing a single representative composition that is then substituted into Eq.~(\ref{eq:kappa_eff}).  At low energies ($E_\gamma \lesssim 0.1\,$MeV), the photoelectric effect dominates and the opacity is strongly composition-dependent, with the $r$-process-rich mid-latitude material exhibiting $\kappa_{\rm eff}$ up to an order of magnitude higher than the equatorial iron-group material due to the $Z^{4\text{--}5}$ scaling of the photoelectric cross section.  In the Compton-dominated regime ($0.2 \lesssim E_\gamma \lesssim 5\,$MeV), where our principal emission lines lie, the opacity varies more weakly with composition ($\kappa_{\rm eff} \sim 0.05$--$0.08\,\mathrm{cm^2\,g^{-1}}$), since the Compton cross section per electron is nearly independent of $Z$.  Above ${\sim}\,5\,$MeV, pair production becomes significant, increasing the opacity for heavier compositions.  The effective opacity $\kappa_{\rm eff}$ is essentially time-independent, because it is dominated by the bulk elemental composition --- stable and long-lived species such as Pt, Xe, Te, He, and Ni --- which is set by nucleosynthesis and does not evolve on the timescales of interest.  Consequently, the evolution of the optical depth (Fig.~\ref{fig:kappa_eff}b) is driven entirely by the $t^{-2}$ decrease in column density from homologous expansion.

To compute the escape probability of a gamma-ray emitted in a given cell, we trace a ray from the cell toward the observer through the 2D spherical grid (Fig.~\ref{fig:geometry}).  Without loss of generality (due to axisymmetry), the observer is placed at azimuthal angle $\phi_{\rm obs} = 0$, i.e.\ in the $rz$-plane, at polar angle $\thetaobs$. For a given observer polar angle $\thetaobs$, the ray from cell $i$ at position $(r_i, \theta_i)$ is parameterized by the path length $s$, with $s = 0$ at the emitting cell.  At each azimuthal sampling angle $\phi_l$, the spherical radius and polar angle along the ray,
\begin{align}
  r(s) &= \sqrt{r_i^2 + s^2 + 2\,r_i\,s\cos\alpha}, \label{eq:ray_r} \\
  \cos\theta(s) &= \frac{r_i\cos\theta_i + s\cos\thetaobs}{r(s)}, \label{eq:ray_theta}
\end{align}
are used to look up the local density and composition from the grid, where $\alpha$ is the angle between the cell's position vector and the observer direction (Eq.~\ref{eq:cosalpha}).  Because the ejecta continue to expand during the photon transit time $s/c$, the effective density at distance $s$ along the ray is corrected to the retarded time $t + s/c$:
\begin{equation}
  \rho^{\rm eff}(s) = \rho(s,t) \left(\frac{t}{t + s/c}\right)^{\!3},
  \label{eq:expansion}
\end{equation}
since $\rho \propto t^{-3}$ for homologous expansion.  At each integration step along the ray, we look up $\rho$ and $X_Z$ from the grid and accumulate the element column densities,
\begin{equation}
  \Sigma_Z = \int_0^{s_{\rm max}} X_Z(s) \, \rho^{\rm eff}(s)\,\ds,
  \label{eq:column}
\end{equation}
for each element $Z$ present in the ejecta.  The energy-dependent optical depth is then
\begin{equation}
  \tau(E_\gamma) = \sum_Z \kappa_Z^{\rm eff}(E_\gamma)\,\Sigma_Z,
  \label{eq:tau_E}
\end{equation}
and the escape probability is $P_{\rm esc}(E_\gamma) = e^{-\tau(E_\gamma)}$.  Figure~\ref{fig:kappa_eff}b shows the radial optical depth $\tau(E_\gamma)$ as a function of time for three representative line energies, $^{133}$Xe at $0.08\,$MeV, $^{132}$I at $0.67\,$MeV, and $^{56}$Co at $3.25\,$MeV, and for two angular directions.  The optical depth drops as $\tau \propto t^{-2}$ due to the homologous expansion of the ejecta.  The ejecta remain opaque to lower energy photons for much longer, regardless of the observer direction, consistent with the higher $\kappa_{\rm eff}$ in the photoelectric regime. 
Because we treat Compton scattering as absorption (Section~\ref{sec:opacity}), a photon either escapes along its initial direction or is removed; we neglect the scattered component entirely.  The time delay introduced by scattering is much shorter than the evolutionary timescale of the ejecta, so temporal smearing is negligible.  The main consequence is an underestimate of the continuum flux between the line features, as Compton down-scattered photons would partially fill in the spectral valleys, and minor asymmetries in the line profiles.  These effects are most relevant at early times ($t \lesssim 5\,$d) and for equatorial observers, where the column density and optical depth are highest (Fig.~\ref{fig:kappa_eff}); along polar and mid-latitude directions, the ejecta become transparent at the principal line energies within the first few days, after which the scattering probability is small and our pure-absorption treatment should be accurate.   A full Monte Carlo gamma-ray transport calculation could refine the spectral shape at early epochs; we defer this to future work. \\ 

Although the ejecta properties ($\rho$, $X_Z$) are axisymmetric and independent of the azimuthal angle $\phi$, the ray geometry from a given 3D position to the observer depends on $\phi$, because the emitting cell and the observer generally do not share the same meridional plane.  To capture this variation, we sample $n_\phi = 60$ uniformly spaced azimuthal angles $\phi_l \in [0, \pi]$; the range $[\pi, 2\pi]$ is not sampled separately because $\cos\alpha$ (Eq.~\ref{eq:cosalpha}) and the column density $\Sigma_Z$ are both symmetric under $\phi \to 2\pi - \phi$, so each sample represents both $\phi_l$ and its mirror image.

\subsection{Doppler shift and emergent spectrum}
\label{sec:spectrum}

The ejecta expand homologously, with each cell $i$ at position $(r_i, \theta_i)$ moving at velocity $v_i = r_i/t$.  The expansion velocities in our model range from $\sim$0.02$c$ in the inner ejecta to $\sim$0.6$c$ in the outermost material, so that relativistic corrections are non-negligible.  Defining $\beta = v/c$ and using $\alpha$, the angle between the cell's velocity vector and the observer direction $\hat{q}$, the line-of-sight velocity component is $\beta_{\rm los} = \beta\cos\alpha$, where
\begin{equation}
  \cos\alpha = \sin\theta_i\,\sin\thetaobs\,\cos\phi_l + \cos\theta_i\,\cos\thetaobs.
  \label{eq:cosalpha}
\end{equation}
Here $\theta_i$ is the polar angle of cell $i$ (from the symmetry axis) and $\phi_l$ is the azimuthal sampling angle.  The relativistic Doppler factor for each cell is
\begin{equation}
  \mathcal{D} = \frac{\sqrt{1 - \beta^2}}{1 - \beta_{\rm los}},
  \label{eq:doppler}
\end{equation}
so that a line emitted at rest-frame energy $E_{\gamma,k}$ is observed at $E_{\rm obs} = \mathcal{D}\,E_{\gamma,k}$.  Because the line-of-sight velocity $\beta_{\rm los}$ depends on the azimuthal angle $\phi_l$ of the emitting cell, and because different radial and vertical positions sample different expansion speeds, each intrinsically narrow line is broadened into a profile whose width is determined by the velocity dispersion of the emitting material.  At the characteristic expansion velocity $v \sim 0.1c$, the lines are broadened by $\Delta E_\gamma / E_\gamma \sim 2v/c \sim 0.2$.
 
The emergent escaping gamma-ray luminosity at observer time $t$ and viewing angle $\theta_{\rm obs}$ is
\begin{equation}
  L^{\rm{esc}}_\gamma(t, \theta_{\rm obs}) = \sum_{i,\,\phi_l} \sum_{A,Z} \sum_k
  \frac{\dot{N}_{A,Z}^{(i)}}{n_\phi} \, I_{\gamma,k} \, \mathcal{D}^{(i)}\,E_{\gamma,k} \, P_{\rm esc}(E_{\gamma,k}) ,
  \label{eq:L_gamma}
\end{equation}
where $P_{\rm esc}(E_{\gamma,k}) = e^{-\tau(E_{\gamma,k})}$ is evaluated at the rest-frame line energy (Eq.~\ref{eq:tau_E}) and $\mathcal{D}^{(i)} E_{\gamma,k}$ is the observer-frame photon energy (Eq.~\ref{eq:doppler}). The Doppler factor $\mathcal{D}$ accounts for the blueshift or redshift of each photon as measured by the observer. The factor $1/n_\phi$ arises because the decay rate $\dot{N}_{A,Z}^{(i)}$ is computed from the full ring mass $\Delta m_i$ (which spans all $2\pi$ in azimuth; see \S\ref{sec:methods}), while the escape probability and Doppler factor vary with the azimuthal position $\phi_l$.
The differential flux $F_{\gamma}\,[\mathrm{erg\,MeV^{-1}\,s^{-1}\,cm^{-2}}]$ at distance $d$ is obtained by histogramming the same contributions into observer-frame energy bins $\Delta E_\gamma$:
\begin{align}
  F_{\gamma}(E_\gamma,t,\theta_{\rm obs})\Delta E_{\gamma}= \frac{ L^{\rm{esc}}_\gamma(t, \theta_{\rm obs})}{4\pi d^2},
  \label{eq:F_gamma}
\end{align}
where $L_\gamma^{\rm esc}(E_\gamma,t,\theta_{\rm obs})$ is the escaping luminosity contributed by all terms in Eq.~(\ref{eq:L_gamma}) whose observed energy $\mathcal{D}^{(i)} E_{\gamma,k}$ falls within $[E_\gamma,\, E_\gamma + \Delta E_\gamma]$.

\begin{figure}
	\centering
	\safefig{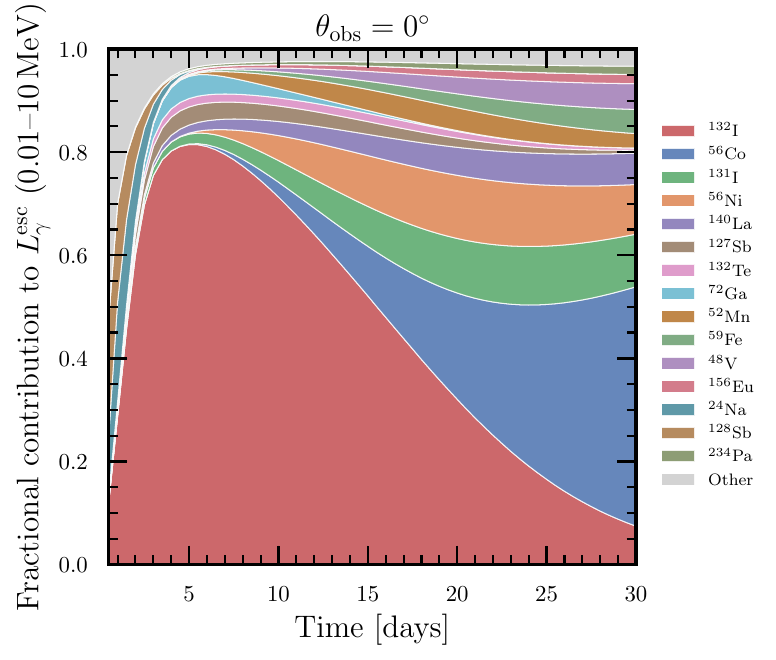}
	\caption{Fractional contribution of individual radioactive isotopes to the total escaping gamma-ray luminosity $L_{\gamma}^{\rm esc}$($0.01$--$10\,\mathrm{MeV}$) as a function of time for the polar viewing angle ($\thetaobs = 0^\circ$).  $^{132}$I dominates between $\sim$1 and $10\,$d, sustained by secular equilibrium with its parent $^{132}$Te ($t_{1/2} = 3.2\,$d).  At late times, $^{56}$Co becomes the primary emitter.}
	\label{fig:L_gamma_stacked}
\end{figure}

\section{RESULTS}
\label{sec:results}

\begin{figure*}
	\centering
	\safefig[width=\textwidth]{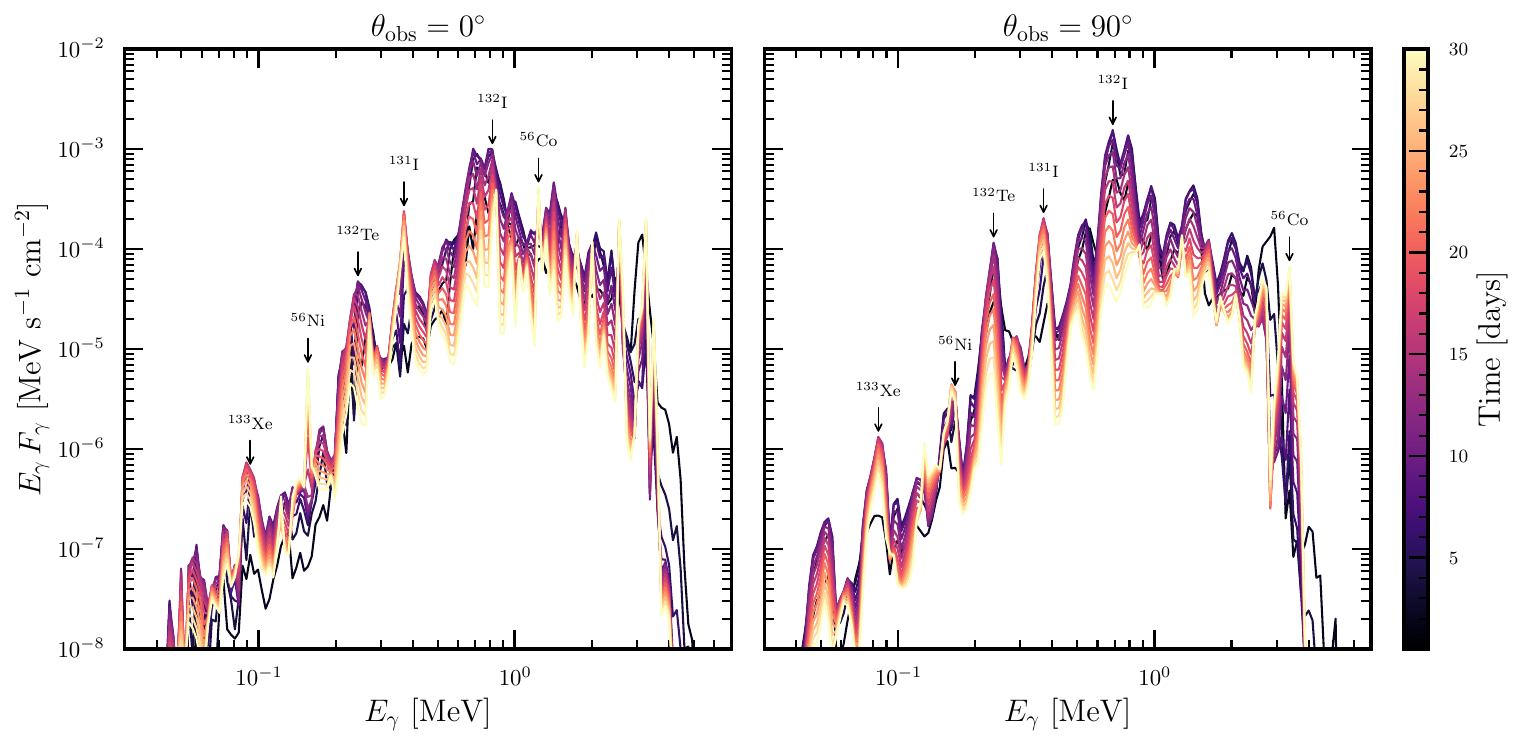}
	\caption{Gamma-ray energy flux (Eq.~\ref{eq:F_gamma}) as a function of time for the polar ($\thetaobs = 0^\circ$, left) and equatorial ($\thetaobs = 90^\circ$, right) viewing angles, for a source at distance $d=1$~Mpc. Each curve corresponds to a different epoch (sampled every $2\,$d from $2$ to $30\,$d), with the color indicating time as shown by the colorbar. The dominant emitting isotopes are labeled at the spectral peaks; an isotope is labeled only when it contributes $\geq 50\,\%$ of the time-integrated fluence in that energy bin (see Fig.~\ref{fig:fluence}) }
	\label{fig:spectrum_isotopes}
\end{figure*}

\subsection{Dominant isotopes and time evolution}
\label{sec:isotopes}

\begin{figure*}
	\centering
	\includegraphics[width=\textwidth]{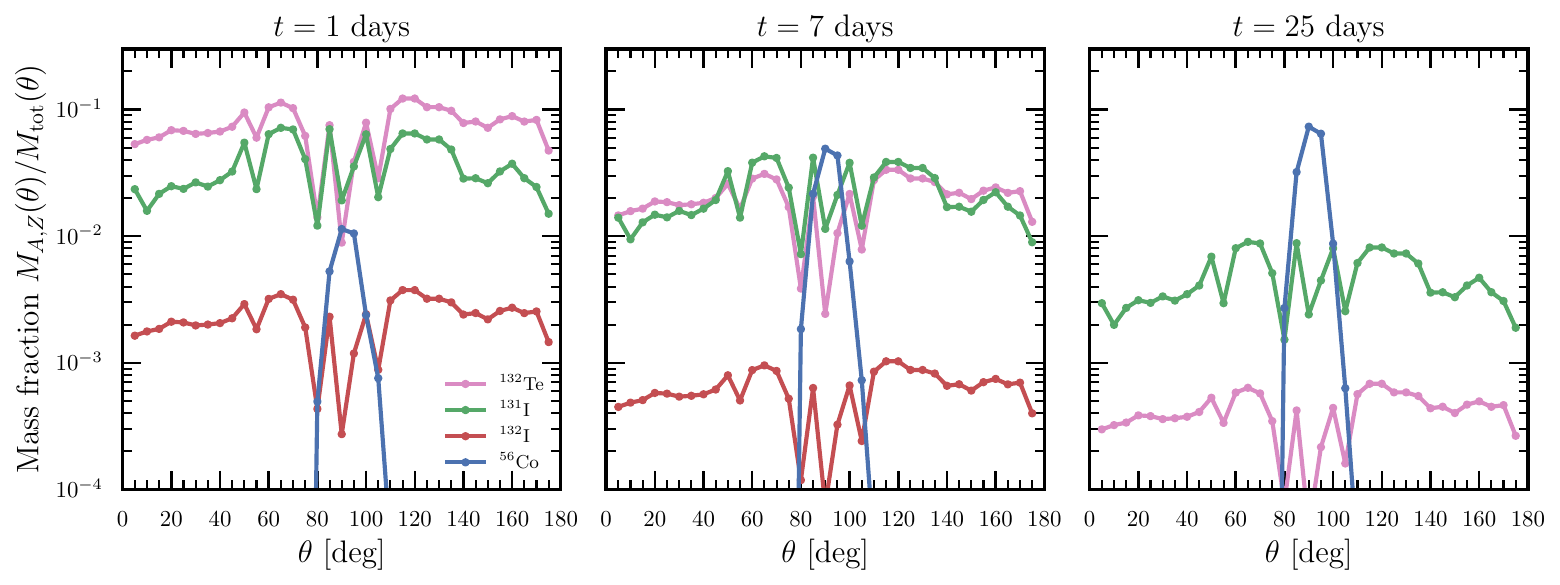}
	\caption{Mass fraction $M_{A,Z}(\theta)/M_{\rm tot}(\theta)$ of the isotopes responsible for the most prominent gamma-ray spectral lines, as a function of polar angle $\theta$, at $t = 1$, $7$, and $25\,$d.  At each angle, the isotope mass is summed over all radial shells and divided by the total ejecta mass in that angular bin.  $^{132}$Te (pink) and its daughter $^{132}$I (red) are broadly distributed across the mid-latitude $r$-process material ($\theta \lesssim 75^\circ$ and $\theta \gtrsim 110^\circ$), while $^{56}$Co (blue, from $^{56}$Ni decay) is confined to the equatorial wedge ($\theta \sim 80^\circ$--$100^\circ$).  $^{131}$I (green) follows a distribution similar to $^{132}$Te but at lower abundance.  The decline of $^{132}$Te and $^{132}$I between $1$ and $25\,$d reflects their radioactive decay ($t_{1/2} = 3.2\,$d and $2.3\,$h, respectively), while $^{56}$Co grows as $^{56}$Ni ($t_{1/2} = 6.1\,$d) decays.}
	\label{fig:isotope_angle}
\end{figure*}

Figure~\ref{fig:L_gamma_stacked} shows the fractional contribution of individual radioactive isotopes to the total escaping gamma-ray luminosity $\Lgamma^{\rm esc}(t)$ for the polar viewing angle ($\thetaobs = 0^\circ$), from $0.5$ to $30\,$d after collapse.  At any given time, a relatively small number of isotopes dominate the total emission, but the identity of the dominant emitters changes markedly with time.

At early times ($t \lesssim 1\,$d), the emission is distributed among several short-lived species: $^{128}$Sb ($t_{1/2} = 9.0\,\mathrm{h}$), $^{129}$Sb ($t_{1/2} = 4.4\,\mathrm{h}$), $^{135}$I ($t_{1/2} = 6.6\,\mathrm{h}$), and $^{24}$Na ($t_{1/2} = 15.0\,\mathrm{h}$), alongside growing contributions from $^{56}$Ni ($t_{1/2} = 6.1\,\mathrm{d}$) and $^{132}$I ($t_{1/2} = 2.3\,\mathrm{h}$).  No single isotope contributes more than $\sim$30\% at these early times.

Between $\sim$1 and $10\,$d, $^{132}$I rises to become the dominant gamma-ray emitter, reaching $\sim$60--70\% of the total luminosity at $t \approx 2$--$3\,$d.  Despite its short half-life ($t_{1/2} = 2.3\,\mathrm{h}$), $^{132}$I persists as a major contributor for several days because it is continuously produced through the $\beta^-$ decay of its parent nucleus $^{132}$Te ($t_{1/2} = 3.2\,\mathrm{d}$).  The half-life of the daughter is much shorter than that of the parent ($t_{1/2}^{\rm I} \ll t_{1/2}^{\rm Te}$), so $^{132}$I reaches secular equilibrium with $^{132}$Te and its effective activity tracks that of $^{132}$Te.  Consequently, the decline of $^{132}$I's contribution beginning at $t \approx 5$--$6\,$d reflects the depletion of the $^{132}$Te reservoir, whose mean lifetime is $ t_{1/2}/\ln 2 \approx 4.6\,$d.  $^{132}$I produces particularly bright gamma-ray lines at $0.668\,\mathrm{MeV}$ (intensity 98.7\%) and $0.773\,\mathrm{MeV}$ (76.5\%), making it the strongest spectral feature at all times between 1 and $10\,$d. 

At late times ($t \gtrsim 10\,$d), $^{56}$Co ($t_{1/2} = 77.2\,\mathrm{d}$, produced by the decay of $^{56}$Ni) emerges as the primary contributor, growing to $\sim$30--40\% by $25\,$d.  $^{52}$Mn ($t_{1/2} = 5.6\,\mathrm{d}$), $^{131}$I ($t_{1/2} = 8.0\,\mathrm{d}$), $^{72}$Ga ($t_{1/2} = 14.1\,\mathrm{h}$), and $^{127}$Sb ($t_{1/2} = 3.9\,\mathrm{d}$) provide secondary contributions at various epochs.

The spatial distributions of the dominant gamma-ray emitters reflect the $\Ye$-dependent nucleosynthesis channels operating in the ejecta.  $^{132}$Te, the parent of $^{132}$I and a second $r$-process peak nucleus, is synthesized mainly in the neutron-rich mid-latitude material ($\Ye \lesssim 0.25$, $\theta <75^\circ$ and $\theta>110^\circ$), where the conditions favour heavy element production.   In contrast, $^{56}$Co --- the dominant late-time emitter --- inherits the spatial distribution of its parent $^{56}$Ni, which is produced in the higher-$\Ye$ equatorial material ($\Ye \gtrsim 0.4$) and is thus confined to a narrow wedge near the equatorial plane.  This spatial distribution directly drives the viewing-angle dependence of the gamma-ray light curve (Fig.~\ref{fig:L_gamma_angle}): the broadly distributed $^{132}$I emission is visible from all directions, while the equatorially confined $^{56}$Co emission escapes preferentially along the polar axis where the overlying column density is lowest.  The angular mass-fraction profiles of these key isotopes at representative epochs are shown in Fig.~\ref{fig:isotope_angle}.

These dominant isotopes are broadly consistent with the predictions of \citep{Chen:2021tob}, who computed gamma-ray spectra for BNS merger ejecta with a range of electron fractions using the XCOM opacity database.  Their Table~1 lists the dominant gamma-ray nuclides for $\Ye = 0.20$ dynamical ejecta -- close to the mean $\Ye \approx 0.24$ of our AIC model -- as $^{132}$I, $^{128}$Sb, $^{133}$I, $^{135}$I, and $^{129}$Sb at $t = 1\,$d, and $^{132}$I, $^{131}$I, $^{132}$Te, $^{127}$Sb, and $^{140}$La at $t = 10\,$d.  While the detailed ranking of sub-dominant species differs -- \citeauthor{Chen:2021tob} use a 1D spherical geometry with a spatially uniform $\Ye$ per run, whereas our 2D calculation captures the continuous variation of $\Ye$ and composition across the ejecta -- the overall set of dominant emitters is in good agreement with our findings.

Similar secular equilibrium effects are likely at work for other parent--daughter pairs in the ejecta (e.g.\ $^{131}$Te$\,\to\,$$^{131}$I), but the $^{132}$Te$\,\to\,$$^{132}$I chain is uniquely prominent because of the large $^{132}$Te yield and the exceptionally bright $^{132}$I gamma-ray spectrum.

\begin{figure*}
	\centering
	\safefig[width=\textwidth]{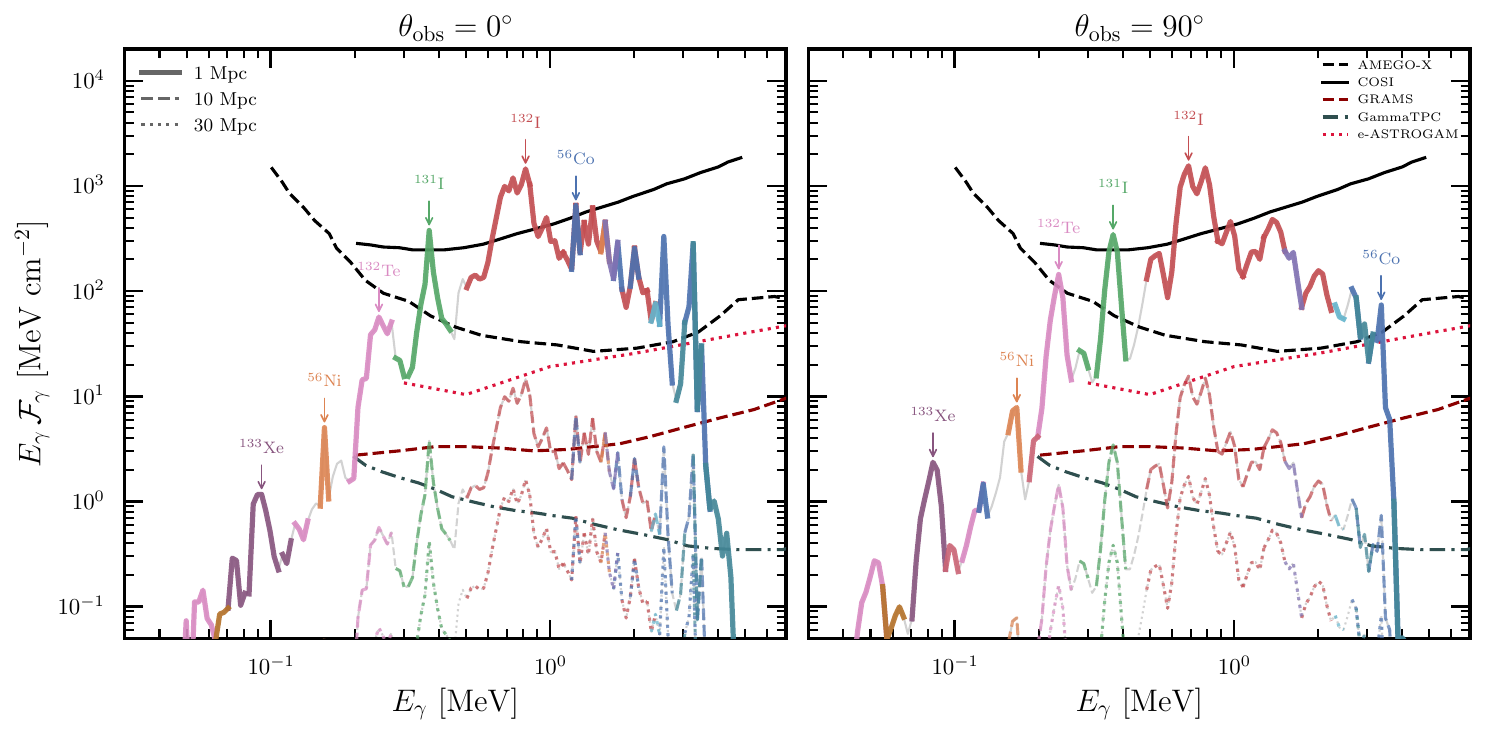}
	\caption{Time-integrated gamma-ray fluence (Eq.~\ref{eq:F_gamma} integrated from $0.5$ to $30\,$d) for the polar ($\thetaobs = 0^\circ$, left) and equatorial ($\thetaobs = 90^\circ$, right) viewing angles. In each energy bin the curve is coloured by the dominant emitting isotope, provided that isotope contributes $\geq 50\,\%$ of the bin fluence; bins without a clear dominant contributor retain the grey background curve.  Each dominant isotope is identified by an arrow label at its spectral peak.  Three source distances are shown with solid (1 Mpc), dashed (10 Mpc), and dotted (30 Mpc) lines. Superimposed are the minimum detectable fluence curves of proposed MeV gamma-ray telescopes, rescaled to a $30\,$d transient as described in Section~\ref{sec:spectra}.  The $r$-process spectral features, in particular the $^{132}$I lines, survive the time integration.}
	\label{fig:fluence}
\end{figure*}

\subsection{Gamma-ray spectra and detectability}
\label{sec:spectra}

Figure~\ref{fig:spectrum_isotopes} shows the time evolution of the observer-frame gamma-ray energy flux for the polar ($\thetaobs = 0^\circ$, left panel) and equatorial ($\thetaobs = 90^\circ$, right panel) viewing angles, for a source at distance $d = 1\,\mathrm{Mpc}$. 

The spectra exhibit well-resolved emission features in the $0.1$--$3\,\mathrm{MeV}$ range from $r$-process isotopes (iodine, tellurium) alongside $^{56}$Co.  The lines are broadened by the Doppler effect arising from the ejecta expansion velocities.  Despite this broadening, the major emission features remain clearly identifiable.  At early times ($t \lesssim 5\,$d), the spectrum is dominated by the bright $^{132}$I lines near $0.7\,\mathrm{MeV}$; at later times, the overall flux decreases as the radioactive inventory is depleted, and $^{56}$Co features become more prominent.  The viewing-angle dependence is modest: the overall spectral shape is similar between the polar and equatorial directions, though the equatorial flux is slightly higher at early times (consistent with Fig.~\ref{fig:L_gamma_angle}).

To assess detectability at larger distances, we compare the time-integrated fluence spectra (Fig.~\ref{fig:fluence}) with the minimum detectable fluence of several proposed MeV gamma-ray telescopes: COSI \citep{Tomsick:2023aue}, AMEGO-X \citep{Caputo:2022xpx}, e-ASTROGAM \citep{e-ASTROGAM:2017pxr}, GammaTPC \citep{Shutt:2025xvc}, and GRAMS \citep{GRAMS:2025ljc}.  Each instrument publishes a $3\sigma$ continuum sensitivity curve, $F_{\rm min}(E_{\gamma}, T_{\rm ref})$, which gives the minimum broadband energy flux (in units of erg\,cm$^{-2}$\,s$^{-1}$) detectable at photon energy $E_{\gamma}$ after an observation of duration $T_{\rm ref}$.  In the background-dominated regime --- which applies at the detection threshold, where the source signal is faint compared to the instrumental and astrophysical backgrounds --- the signal-to-noise ratio grows as $\sqrt{T_{\rm obs}}$, so the minimum detectable flux scales as $F_{\rm min} \propto 1/\sqrt{T_{\rm obs}}$.  To convert the published flux sensitivity into a minimum detectable \emph{fluence} for a transient lasting $T_{\rm obs} \approx 30\,$d, we use $\mathcal{F}_{\rm min}(E_{\gamma}) = F_{\rm min}(E_{\gamma}, T_{\rm ref})\,\sqrt{T_{\rm ref}\,T_{\rm obs}}\,,$
where $T_{\rm ref}$ is the reference observation time adopted in the published curve and ranges from $10^6\,$s for e-ASTROGAM to $\sim 6 \times 10^7\,$s for COSI (2-yr survey).  These results demonstrate that the gamma-ray signature of an AIC event could be detectable at cosmologically relevant distances, far beyond the Local Group.

A potential concern is that the individual gamma-ray lines, although bright, may be washed out by the long integration times ($\sim$10$^6$\,s) required by even the most sensitive detectors.  As different radioactive isotopes take turns to dominate the emission spectrum at successive epochs (Fig.~\ref{fig:L_gamma_stacked}), one might expect the time-integrated signal to become a featureless blend.

Figure~\ref{fig:fluence} shows the time-integrated gamma-ray spectral fluence $\mathcal{F} [\mathrm{MeV\,cm^{-2}}]$, obtained by integrating the instantaneous spectra from $0.5$ to $30\,$d, for the same two viewing angles and three distances as in Fig.~\ref{fig:spectrum_isotopes}. Energy bins are coloured by the dominant emitting isotope whenever it contributes $\geq 50\,\%$ of the bin fluence, and each dominant isotope is identified by an arrow label at its spectral peak.

Remarkably, the spectral features survive the time integration: the $^{132}$I lines at $0.67$ and $0.77\,\mathrm{MeV}$ remain the strongest features in the fluence spectrum, followed by $^{131}$I ($0.36\,\mathrm{MeV}$), $^{132}$Te ($0.23\,\mathrm{MeV}$), $^{133}$Xe ($0.08\,\mathrm{MeV}$), $^{56}$Ni ($0.16\,\mathrm{MeV}$), and $^{56}$Co ($2.60\,\mathrm{MeV}$).  These features are characteristic products of the $r$-process -- iodine, tellurium, xenon-- and provide unambiguous spectral fingerprints of neutron-rich nucleosynthesis.  Their persistence in the time-integrated signal is crucial for observational prospects, since the long integration times ($\sim 10^6\,$s) required by current and planned MeV instruments would not wash out the identifying spectral signatures. The energy resolution of the instruments considered -- $\lesssim 1\,\%$ for COSI \citep{Tomsick:2023aue}, $\sim 5\,\%$ FWHM at $1\,\mathrm{MeV}$ for AMEGO-X \citep{Caputo:2022xpx}, and $\sim 5\,\%$ for e-ASTROGAM \citep{e-ASTROGAM:2017pxr} in the Compton regime -- is well below the intrinsic Doppler broadening of the lines ($\Delta E_\gamma / E_\gamma \sim 0.2$ at $v \sim 0.1\,c$), so the spectral features are fully resolved by the detectors.

\subsection{Element abundance vs.\ gamma-ray energy budget}
\label{sec:scatter}

\begin{figure}
  \centering
  \safefig{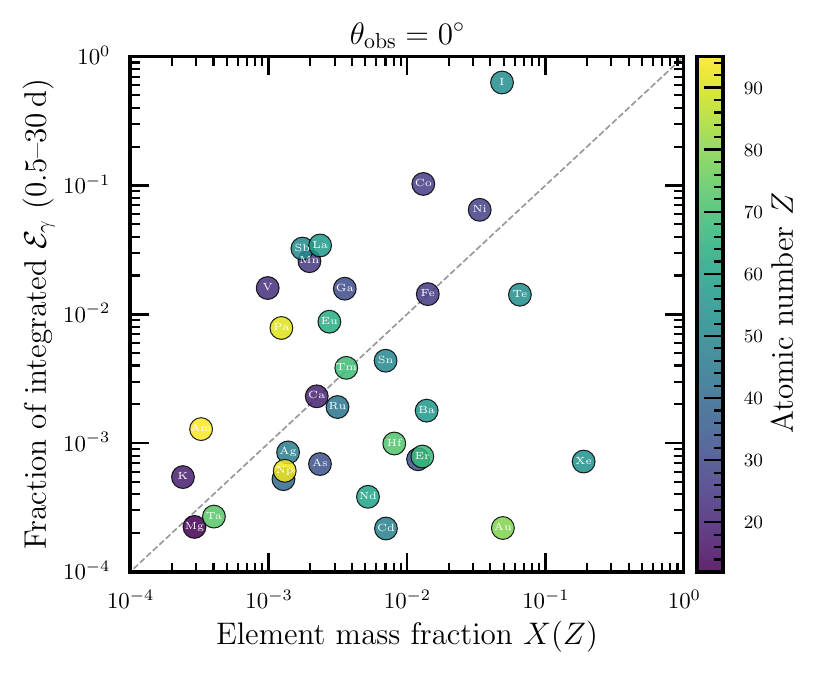}
  \caption{Elemental mass fraction $X(Z)$ vs.\ fraction of time-integrated gamma-ray energy ($0.5$--$30\,$d) for each element, at $\thetaobs = 0^\circ$.  Points are coloured by atomic number $Z$.  The dashed line marks equal fractions.  Iodine ($Z = 53$) dominates the gamma-ray energy budget with $\sim$40\% despite comprising only $\sim$5\% of the ejecta mass.}
  \label{fig:scatter}
\end{figure}

Figure~\ref{fig:scatter} relates the elemental mass fraction $X(Z)$ of each element in the ejecta to its fractional contribution to the total time-integrated gamma-ray energy over the $0.5$--$30\,$d interval, for the polar viewing angle.  The mass fractions are computed from the \texttt{SkyNet} composition at the end of the nucleosynthesis calculation ($t \sim 30\,$d), summed over all isotopes of each element and weighted by solid angle. Because the mass fractions refer to the final composition, they include the stable daughter products of earlier radioactive decays (e.g.\ $^{132}$Xe from $^{132}$I $\to$ $^{132}$Xe); the gamma-ray energy, on the other hand, was produced by the radioactive parents during the preceding $30\,$d.  This is the natural comparison: the final $X(Z)$ is the well-defined nucleosynthetic yield of each element, while $\mathcal{E}_\gamma(Z)$ measures the cumulative diagnostic power of its radioactive isotopes. The dashed diagonal line marks $X(Z) = \mathcal{E}_\gamma(Z)/\mathcal{E}_{\gamma,\rm tot}$, i.e.\ where the gamma-ray energy fraction would equal the mass fraction.

Iodine ($Z = 53$) stands out as the most important gamma-ray emitter: despite a mass fraction of only $X \sim 5\%$, it accounts for $\sim$40\% of the total escaping gamma-ray energy.  This outsized contribution arises from the bright $^{132}$I lines and the favourable combination of half-life and escape fraction.  Other elements lying well above the diagonal -- including Ni ($Z = 28$), Co ($Z = 27$), Sb ($Z = 51$), Mn ($Z = 25$), and Ga ($Z = 31$) -- are similarly ``gamma-ray loud'': they contribute disproportionately to the gamma-ray output relative to their mass.  Conversely, elements with large mass fractions but modest gamma-ray contributions, such as Te ($Z = 52$) and Xe ($Z = 54$), fall closer to or below the diagonal: although $^{132}$Te produces recognisable spectral features (Fig.~\ref{fig:fluence}), the bulk of the tellurium mass resides in stable or weakly emitting isotopes, so the element-integrated gamma-ray fraction remains modest relative to its total abundance.

This figure illustrates a key feature of radioactive gamma-ray diagnostics: the gamma-ray spectrum is not a simple tracer of total elemental abundance.  Instead, it preferentially highlights those elements whose radioactive isotopes have (i) half-lives comparable to the ejecta transparency timescale ($\sim$days), (ii) high gamma-ray line energies and branching ratios, and (iii) lines in the Compton-dominated regime ($E_\gamma \gtrsim 0.2\,\mathrm{MeV}$) where the escape probability is highest.

\subsection{Viewing-angle dependence of the gamma-ray light curve}
\label{sec:angle_dependence}

\begin{figure}
  \centering
  \safefig{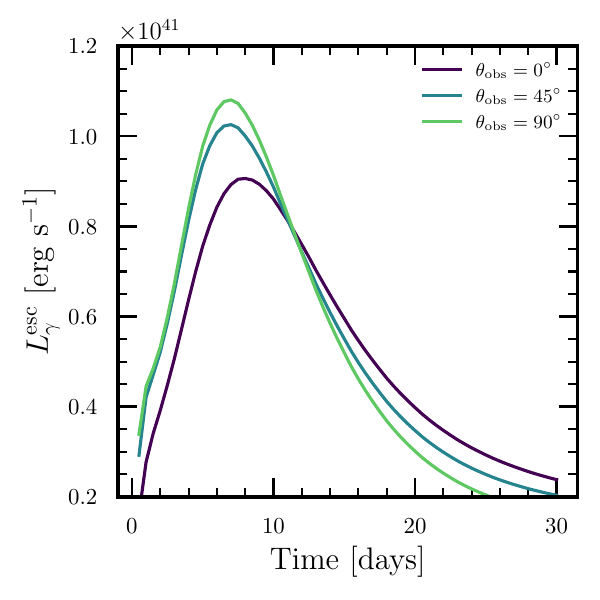}
  \caption{Total escaping gamma-ray luminosity $\Lgamma^{\rm esc}(t)$ for three observer viewing angles: $\thetaobs = 0^\circ$ (polar), $45^\circ$ (intermediate), and $90^\circ$ (equatorial).  The light-curve crossover at $t \sim 5$--$10\,$d reflects the transition from $^{132}$I-dominated (equatorially brighter) to $^{56}$Co-dominated (polar brighter) emission, as described in the text.}
  \label{fig:L_gamma_angle}
\end{figure}

\begin{figure*}
	\centering
	\safefig[width=\textwidth]{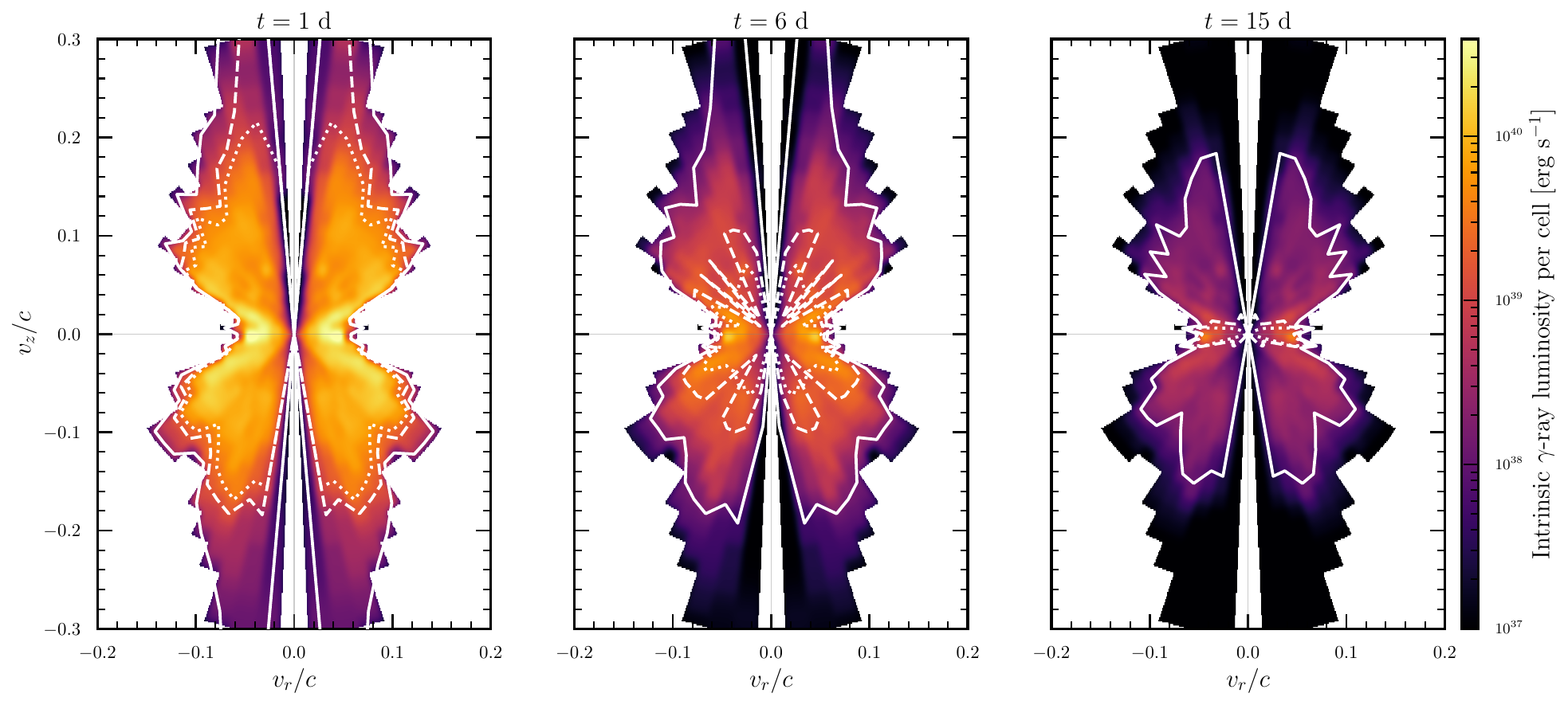}
	\caption{Intrinsic gamma-ray luminosity per cell in velocity space at $t = 1$, $6$, and $15\,$d after collapse.  White curves show the gamma-ray photosphere ($\tau_\gamma = 1$) at $0.08\,\mathrm{MeV}$ ($^{133}$Xe, solid), $0.67\,\mathrm{MeV}$ ($^{132}$I, dashed), and $3.25\,\mathrm{MeV}$ ($^{56}$Co, dotted).  At $t = 1\,$d the photosphere is elongated toward the poles; by $t = 15\,$d it has become oblate, compressed to the equatorial region where the column density is highest.  This evolution is consistent with the preferentially polar escape of gamma-rays at late times (Fig.~\ref{fig:L_gamma_angle}).}
	\label{fig:ejecta_2D_lum}
\end{figure*}

Figure~\ref{fig:L_gamma_angle} shows the total escaping gamma-ray luminosity $\Lgamma^{\rm esc}(t)$ for three observer viewing angles.  At early times ($t \lesssim 3$--$5\,$d), the equatorial observer ($\thetaobs = 90^\circ$) sees a brighter and earlier-peaking light curve than the polar observer ($\thetaobs = 0^\circ$); at later times ($t \gtrsim 10\,$d), the situation reverses.  The origin of this crossover can be understood from Fig.~\ref{fig:ejecta_2D_lum}.  At $t \lesssim 10\,$d the emission is dominated by $^{132}$I (Fig.~\ref{fig:L_gamma_stacked}), whose escaping luminosity is slightly higher toward the equator.  At $t \gtrsim 10\,$d, $^{56}$Co takes over as the dominant emitter.  Because $^{56}$Co is confined to the equatorial region, its photons escape more easily toward the pole, where the column density is lowest; along the equatorial plane, the gamma-ray photosphere at $\sim$\,MeV energies still extends to the outer ejecta even at these late epochs (Fig.~\ref{fig:ejecta_2D_lum}), suppressing the equatorial escape.  This shift in the dominant emitter -- from the broadly distributed $^{132}$I to the equatorially confined $^{56}$Co -- drives the light-curve crossover.

The spatial structure of the gamma-ray emission and opacity is illustrated in Fig.~\ref{fig:ejecta_2D_lum}, which shows the intrinsic gamma-ray luminosity per cell in velocity space at $t = 1$, $6$, and $15\,$d after collapse.  Overlaid are contours of radial gamma-ray optical depth $\tau_\gamma = 1$ (integrated outward from each cell along the radial direction) at three photon energies corresponding to the principal emission lines: $0.08\,\mathrm{MeV}$ ($^{133}$Xe, solid), $0.67\,\mathrm{MeV}$ ($^{132}$I, dashed), and $3.25\,\mathrm{MeV}$ ($^{56}$Co, dotted).  These radial photospheres indicate where the ejecta transition from optically thick to thin along the expansion direction; the effective photosphere seen by an observer at a given viewing angle, defined by integrating $\tau$ along the line of sight, would differ in shape due to the aspherical ejecta geometry. Nevertheless, the radial photospheres provide a useful visualisation of the spatial extent of the opaque region and its energy-dependent recession over time.


At $t = 1\,$d, the ejecta are largely opaque at $0.08\,\mathrm{MeV}$ (photoelectric regime) and $0.67\,\mathrm{MeV}$ (the dominant $^{132}$I line energy), with the photospheres extending to the outermost ejecta layers and elongated toward the polar axis.  The $3.25\,\mathrm{MeV}$ photosphere ($^{56}$Co) sits deeper, reflecting the lower Compton cross section at higher energies (cf.\ Fig.~\ref{fig:kappa_eff}a).  As time advances and the ejecta expand, all photospheres recede inward.  The shape evolves from prolate at early times -- when the polar ejecta are still optically thick -- to oblate by $t = 15\,$d, when the photosphere is compressed to the equatorial region where the column density remains highest.  This flattening is consistent with the viewing-angle dependence of the light curve (Fig.~\ref{fig:L_gamma_angle}): at late times, photons escape more easily along the lower-density polar axis while the equatorial direction remains opaque.  The mild north--south asymmetry in the photosphere reflects slight differences between the two hemispheres already present in the underlying GRMHD simulation.

The persistence of a large equatorial optical depth is also consistent with the results of our companion paper \citep{Pitik:2026bjm}, where the kilonova optical/infrared luminosity was found to originate predominantly from the equatorial region.  That emission is powered by the thermalization of gamma-ray photons in the densest parts of the ejecta -- precisely the regions where the photosphere in Fig.~\ref{fig:ejecta_2D_lum} remains at large radii, indicating efficient absorption and energy deposition.

\section{DISCUSSION}
\label{sec:discussion}

\subsection{Comparison with binary neutron star mergers}

Several studies have predicted the early-time ($t \lesssim 30\,$d) gamma-ray emission from BNS merger ejecta. \citep{Li:2019gamma} presented a semi-analytic model predicting a peak gamma-ray luminosity of $\sim 2 \times 10^{41}\,\mathrm{erg\,s^{-1}}$ at $\sim$1.2\,d, detectable to $\lesssim 12\,\mathrm{Mpc}$.  \citep{Korobkin:2019uxw} performed 3D Monte Carlo gamma-ray transport with the Maverick code for a two-component model ($M_{\rm dyn} = 0.0065\,\Msun$ at $v = 0.2c$, $M_{\rm wind} = 0.03\,\Msun$ at $v = 0.08c$), identifying specific emitting isotopes (e.g.\ $^{128}$Sb, $^{131}$I, $^{133}$Xe) as a function of electron fraction and fission model, and estimating detectability to $\sim 3$\,Mpc with AMEGO/COSI detectors.  \citep{Chen:2021tob} computed gamma-ray spectra using NIST-XCOM composition-dependent opacities for a systematic grid of electron fractions ($\Ye = 0.05$--$0.40$), confirming that the $^{132}$Te$\,\to\,$$^{132}$I chain dominates for $\Ye \lesssim 0.25$, while lighter species ($^{72}$Ga, $^{77}$Ge) dominate at $\Ye \gtrsim 0.30$.

The gamma-ray signatures of our AIC model share the same $r$-process features: the $^{132}$Te$\,\to\,$$^{132}$I chain is the brightest emitter during the first $\sim$10\,d, as in BNS models with comparable electron fractions.  An important compositional difference, however, is the presence of significant $^{56}$Ni and $^{56}$Co in the AIC ejecta.  None of the BNS gamma-ray calculations cited above produce iron-peak nuclei -- even at $\Ye = 0.40$, \citep{Chen:2021tob} find the dominant species near the first $r$-process peak ($A \sim 70$--$80$) rather than the iron group ($A \sim 56$).  \citep{Jacobi:2025eak} showed that $^{56}$Ni can be synthesized in the BNS neutrino-driven wind via alpha-rich freeze-out, but only when the electron fraction reaches $\Ye \gtrsim 0.48$ and the NS remnant is long-lived (not promptly collapsing to a black hole), yielding $\sim 10^{-3}\,\Msun$ of $^{56}$Ni.  In contrast, our AIC ejecta ($\langle \Ye \rangle \sim 0.24$, with equatorial material reaching $\Ye \gtrsim 0.3$) produce both $r$-process isotopes and iron-peak species simultaneously.

Quantitative differences also arise from the larger ejecta mass (${\sim} 0.18\,\Msun$ vs.\ ${\sim} 0.005$--$0.03\,\Msun$ for typical BNS models).  The larger mass produces a proportionally higher intrinsic gamma-ray luminosity, but also extends the optically thick phase: the escaping luminosity $L_\gamma^{\rm esc} \propto L_\gamma^{\rm int} \times f_{\rm esc}$ reflects a competition between the source strength and the absorption, which might partially offset each other.  
The highly aspherical geometry of the AIC ejecta -- concentrated at mid-latitudes rather than distributed in the equatorial plane -- introduces a distinct viewing-angle dependence. Table~\ref{tab:comparison} summarizes the key differences.  We caution that the AIC values in Table~\ref{tab:comparison} correspond to a single model --- the $B_{\rm pol} = 10^{12}\,$G case --- which is the only model in the current 2D simulation suite that produces strong $r$-process nucleosynthesis.  As discussed in Section~\ref{sec:simulation}, this strong initial field serves as a proxy for MRI/dynamo amplification that cannot be resolved in axisymmetry, and 3D
simulations with weaker seed fields may yield similar outcomes \citep{Combi:2025yvs}.  Nevertheless, the quoted $\langle \Ye \rangle \sim 0.24$ is quite representative of the bulk of the ejecta, with
${\sim}80\%$ of the total ejected mass having $\Ye \leq 0.25$ \citep{Pitik:2026bjm}. 

In a multi-messenger context, several observables can help distinguish AIC from BNS events.  If the transient occurs within the sensitivity volume of gravitational-wave detectors (LIGO/Virgo/KAGRA), the GW signal itself is diagnostic: a BNS merger produces a characteristic inspiral chirp, whereas an AIC generates a short burst from the collapse and subsequent proto-neutron star oscillations \citep{Abdikamalov:2009aq, 2023MNRAS.525.6359L}.  Most decisively, a coincident GW detection with the inspiral chirp signature characteristic of a compact binary merger would unambiguously rule out the WD collapse channel.  Optical and near-infrared observations would detect the associated kilonova emission \citep{Pitik:2026bjm}, constraining the ejecta mass and opacity.  The gamma-ray lines, in turn, would directly identify the nucleosynthesis products, breaking degeneracies inherent in broadband light-curve fitting.  In particular, the simultaneous detection of $r$-process gamma-ray lines from $^{132}$I, $^{132}$Te, $^{133}$Xe, alongside the iron-peak lines of $^{56}$Co and $^{56}$Ni would favour an AIC origin over a BNS merger, since typical BNS dynamical ejecta lack $^{56}$Co.  

\begin{table}
  \centering
  \caption{Comparison of gamma-ray emission properties between a strongly magnetized and rotating AIC and BNS merger ejecta.  The AIC values are from~\citep{Pitik:2026bjm}; BNS values are representative ranges from the literature \citep{Hotokezaka:2015cma, Korobkin:2019uxw, Chen:2021tob, Jacobi:2025eak}.}
  \label{tab:comparison}
  \begin{tabular}{lcc}
    \hline
    Property & AIC & BNS merger \\
    \hline
    $M_{\rm ej}$ [$\Msun$] & $0.18$ & $0.005$--$0.03$ \\
    $v_{\rm ej}$ [$c$] & $0.02$--$0.6$ & $0.08$--$0.3$ \\
    $\langle \Ye \rangle$ & $\sim 0.24$ & $0.05$--$0.45$ \\
    Dominant $\gamma$-emitter & $^{132}$I & $^{132}$I$^a$ \\
    \quad (1--10\,d) & & \\
    $r$-process lines & Yes & Yes \\
    $^{56}$Ni/$^{56}$Co lines & Sub-dominant & Absent$^b$ \\
    \hline
    \multicolumn{3}{l}{\footnotesize $^a$ For $\Ye \lesssim 0.25$; lighter species ($^{72}$Ga, $^{77}$Ge)}\\
    \multicolumn{3}{l}{\footnotesize \phantom{$^a$} dominate at higher $\Ye$ \citep{Chen:2021tob}.}\\
    \multicolumn{3}{l}{\footnotesize $^b$ In dynamical ejecta; $\sim 10^{-3}\,\Msun$ possible in}\\
    \multicolumn{3}{l}{\footnotesize \phantom{$^b$} high-$\Ye$ wind if remnant is long-lived}\\
    \multicolumn{3}{l}{\footnotesize \phantom{$^b$} \citep{Jacobi:2025eak}.}
  \end{tabular}
\end{table}

\subsection{Uncertainties}
\label{sec:uncertainties}

Several sources of uncertainty affect our predicted gamma-ray fluxes: \\

1) \emph{Hydrodynamics and neutrino transport}: the total ejecta mass ($M_{\rm ej} \sim 0.18\,\Msun$), $\Ye$ distribution, and velocity profile depend on the specific AIC model (magnetic field strength, rotation rate, WD mass).  The gamma-ray luminosity scales approximately linearly with $M_{\rm ej}$, while the composition (and thus the dominant emitting isotopes) is sensitive to the $\Ye$ distribution, which in turn is governed by the neutrino irradiation of the ejecta.  The treatment of neutrino transport --- in particular the neutrino luminosities, spectra, and the duration over which the ejecta are exposed to the neutrino field --- directly controls the degree of neutron-richness and therefore which $r$-process isotopes are synthesized.  Different initial conditions for the magnetic field and rotation, as well as different approximations for the neutrino transport, would result in different amounts of ejecta with distinct compositions, which could modify our conclusions regarding both the detectability and the spectroscopic features of such an event.

2) \emph{Nuclear physics}: the reaction rates for exotic, neutron-rich nuclei far from stability are based on theoretical models (e.g.\ FRDM masses, Hauser--Feshbach cross sections) that carry uncertainties of factors of $\sim$2--3 in the individual isotope yields \citep{Lippuner:2017tyn, Moller:2015fba}. The production yields of our dominant emitters ($^{132}$I, $^{131}$I) are therefore subject to these uncertainties.  However, their decay properties -- half-lives, gamma-ray energies, and branching ratios -- are experimentally measured to high precision, so the predicted emission per decay is not a source of uncertainty; the nuclear-physics uncertainty enters solely through the production yields.

3) \emph{Opacity}: our XCOM-based opacity model includes composition-dependent Compton, photoelectric, and pair-production cross sections for all elements; the remaining uncertainty is the neutral-atom approximation inherent in XCOM, which might be not negligible at low gamma-ray energies (${\sim} 0.01{-}0.1$~MeV). Furthermore, as discussed in Section~\ref{sec:opacity}, our ray-tracing treats Compton scattering as effective absorption.  The primary consequence is an underestimate of the continuum flux between emission lines; the line fluxes and total luminosity are robust.  A full Monte Carlo gamma-ray transport calculation could quantify the scattered continuum and is deferred to future work.

\section{CONCLUSIONS}
\label{sec:conclusions}

We have presented the first predictions of the gamma-ray line emission from $r$-process nuclei produced in the accretion-induced collapse of a magnetized white dwarf.  Our calculation uses time-dependent isotopic abundances from the nuclear reaction network \texttt{SkyNet} with composition-dependent opacities from the NIST-XCOM database, and employs energy-dependent ray-tracing through the two-dimensional ejecta structure.  Our main findings are:

\begin{enumerate}
  \item The escaping gamma-ray luminosity from the AIC ejecta is dominated at any given time by a small number of radioactive isotopes.  Between $\sim$1 and $10\,$d, $^{132}$I accounts for $\sim$60--70\% of the total gamma-ray output, sustained by secular equilibrium with its parent $^{132}$Te ($t_{1/2} = 3.2\,$d).  At later times ($t \gtrsim 20\,$d), $^{56}$Co ($t_{1/2} = 77\,$d) becomes the primary emitter.

  \item The gamma-ray spectra exhibit well-resolved emission features from $r$-process--$^{132}$I, $^{131}$I, $^{132}$Te, $^{133}$Xe,  and others--that provide unambiguous signatures of neutron-rich nucleosynthesis.

  \item At $d = 1\,\mathrm{Mpc}$, the gamma-ray signal is well above the sensitivity of all instruments considered (COSI, AMEGO-X, e-ASTROGAM, GammaTPC, GRAMS).  At $d = 10\,\mathrm{Mpc}$, the brightest $r$-process lines remain detectable by GammaTPC and GRAMS, and are marginally above the e-ASTROGAM threshold.  At $d = 30\,\mathrm{Mpc}$, the signal approaches the GammaTPC and GRAMS sensitivity levels.

  \item The $r$-process spectral features survive time integration over the full $0.5$--$30\,$d observation window (Fig.~\ref{fig:fluence}), demonstrating that the long exposure times required by gamma-ray telescopes do not wash out the isotopic signatures.

  \item The gamma-ray energy budget is dominated by iodine ($Z = 53$), which contributes $\sim$40\% of the total escaping gamma-ray energy despite comprising only $\sim$5\% of the ejecta mass (Fig.~\ref{fig:scatter}).  The total time-integrated escaping gamma-ray energy is $\mathcal{E}_{\gamma,\rm esc} \sim  10^{47}\,\mathrm{erg}$.

  \item The AIC gamma-ray spectrum shares the $r$-process features (dominated by the $^{132}$Te\,$\to$\,$^{132}$I chain) predicted for BNS merger ejecta, but additionally contains $^{56}$Co lines that are absent in typical BNS dynamical ejecta.  The simultaneous detection of $r$-process and iron-peak gamma-ray lines would favour an AIC origin.  In a multi-messenger setting, the gravitational-wave signal provides the most decisive discriminant: a BNS merger produces an inspiral chirp, whereas an AIC generates a collapse-bounce signal, unambiguously identifying the progenitor channel.
\end{enumerate}

These results establish gamma-ray observations as a powerful and direct diagnostic of $r$-process nucleosynthesis in AIC events.  The detection horizon extends to $\sim 10\,\mathrm{Mpc}$ for the most sensitive proposed MeV telescopes (GammaTPC, GRAMS), well beyond the Local Group, making AIC events potential targets for next-generation gamma-ray missions.

\begin{acknowledgments}

TP acknowledges support from NSF Grant PHY-2020275 (Network for Neutrinos, Nuclear Astrophysics, and Symmetries (N3AS)).
YZQ was supported in part by the US Department of Energy under grant DE-FG02-87ER40328.
DR acknowledges support from the Sloan Foundation, from the Department of Energy, Office of Science, Division of Nuclear Physics under Awards Number DOE DE-SC0021177 and DE-SC0024388, from the National Science Foundation under Grants No. PHY-2020275, PHY-2407681, and PHY-2512802.
DK is supported in part by the U.S. Department of Energy, Office of Science, Office of Nuclear Physics, DE-AC02-05CH11231, DE-SC0004658, and DE-SC0024388, and by a grant from the Simons Foundation (622817DK).
The numerical simulations were performed on Perlmutter using NERSC award ERCAP0031370. This research used resources of the National Energy Research Scientific Computing Center, a DOE Office of Science User Facility supported by the Office of Science of the U.S.~Department of Energy under Contract No.~DE-AC02-05CH11231.

\end{acknowledgments}

\bibliography{refs}

@ARTICLE{1986PrPNP..17..249N,
       author = {{Nomoto}, K.},
        title = "{The fate of accreting white dwarfs: Type I supernovae vs. collapse.}",
      journal = {Progress in Particle and Nuclear Physics},
         year = 1986,
        month = jan,
       volume = {17},
        pages = {249-266},
          doi = {10.1016/0146-6410(86)90020-7},
       adsurl = {https://ui.adsabs.harvard.edu/abs/1986PrPNP..17..249N}
}

@ARTICLE{1991ApJ...367L..19N,
       author = {{Nomoto}, Ken'ichi and {Kondo}, Yoji},
        title = "{Conditions for Accretion-induced Collapse of White Dwarfs}",
      journal = {\apjl},
         year = 1991,
        month = jan,
       volume = {367},
        pages = {L19},
          doi = {10.1086/185922},
       adsurl = {https://ui.adsabs.harvard.edu/abs/1991ApJ...367L..19N}
}

@article{Abdikamalov:2009aq,
    author = "Abdikamalov, E. B. and Ott, C. D. and Rezzolla, L. and Dessart, L. and Dimmelmeier, H. and Marek, A. and Janka, H. -T.",
    title = "{Axisymmetric General Relativistic Simulations of the Accretion-Induced Collapse of White Dwarfs}",
    eprint = "0910.2703",
    archivePrefix = "arXiv",
    primaryClass = "astro-ph.HE",
    doi = "10.1103/PhysRevD.81.044012",
    journal = "Phys. Rev. D",
    volume = "81",
    pages = "044012",
    year = "2010"
}

@ARTICLE{2023MNRAS.525.6359L,
       author = {{Longo Micchi}, Lu{\'\i}s Felipe and {Radice}, David and {Chirenti}, Cecilia},
        title = "{Multimessenger emission from the accretion-induced collapse of white dwarfs}",
      journal = {\mnras},
         year = 2023,
        month = nov,
       volume = {525},
       number = {4},
        pages = {6359-6376},
          doi = {10.1093/mnras/stad2420},
archivePrefix = {arXiv},
       eprint = {2306.04711},
 primaryClass = {astro-ph.HE},
       adsurl = {https://ui.adsabs.harvard.edu/abs/2023MNRAS.525.6359L}
}

@article{Cheong:2024hrd,
    author = "Cheong, Patrick Chi-Kit and Pitik, Tetyana and Longo Micchi, Lu{\'\i}s Felipe and Radice, David",
    title = "{Gamma-Ray Bursts and Kilonovae from the Accretion-induced Collapse of White Dwarfs}",
    eprint = "2410.10938",
    archivePrefix = "arXiv",
    primaryClass = "astro-ph.HE",
    doi = "10.3847/2041-8213/ada1cc",
    journal = "Astrophys. J. Lett.",
    volume = "978",
    number = "2",
    pages = "L38",
    year = "2025"
}

@article{Kuroda:2025iyj,
    author = "Kuroda, Takami and Kawaguchi, Kyohei and Shibata, Masaru",
    title = "{Collapse of rotating white dwarfs and multimessenger signals}",
    eprint = "2503.17082",
    archivePrefix = "arXiv",
    primaryClass = "astro-ph.HE",
    doi = "10.1093/mnras/staf1065",
    journal = "Mon. Not. Roy. Astron. Soc.",
    volume = "541",
    number = "2",
    pages = "1649--1669",
    year = "2025"
}

@article{Dessart:2006pe,
    author = "Dessart, L. and Burrows, A. and Ott, C. D. and Livne, E. and Yoon, S.-C. and Langer, N.",
    title = "{Multidimensional Simulations of the Accretion-Induced Collapse of White Dwarfs to Neutron Stars}",
    doi = "10.1086/503626",
    journal = "Astrophys. J.",
    volume = "644",
    pages = "1063--1084",
    year = "2006"
}

@ARTICLE{1999ApJ...516..892F,
       author = {{Fryer}, Chris and {Benz}, Willy and {Herant}, Marc and {Colgate}, Stirling A.},
        title = "{What Can the Accretion-induced Collapse of White Dwarfs Really Explain?}",
      journal = {\apj},
         year = 1999,
        month = may,
       volume = {516},
       number = {2},
        pages = {892-899},
          doi = {10.1086/307119},
archivePrefix = {arXiv},
       eprint = {astro-ph/9812058},
 primaryClass = {astro-ph},
       adsurl = {https://ui.adsabs.harvard.edu/abs/1999ApJ...516..892F}
}

@ARTICLE{2007MNRAS.380..933Y,
       author = {{Yoon}, S. -C. and {Podsiadlowski}, Ph. and {Rosswog}, S.},
        title = "{Remnant evolution after a carbon-oxygen white dwarf merger}",
      journal = {\mnras},
         year = 2007,
        month = sep,
       volume = {380},
       number = {3},
        pages = {933-948},
          doi = {10.1111/j.1365-2966.2007.12161.x},
archivePrefix = {arXiv},
       eprint = {0704.0297},
 primaryClass = {astro-ph},
       adsurl = {https://ui.adsabs.harvard.edu/abs/2007MNRAS.380..933Y}
}

@ARTICLE{2009ApJ...692..324S,
       author = {{Shen}, Ken J. and {Bildsten}, Lars},
        title = "{The Effect of Composition on Nova Ignitions}",
      journal = {\apj},
         year = 2009,
        month = feb,
       volume = {692},
       number = {1},
        pages = {324-334},
          doi = {10.1088/0004-637X/692/1/324},
archivePrefix = {arXiv},
       eprint = {0805.2160},
 primaryClass = {astro-ph},
       adsurl = {https://ui.adsabs.harvard.edu/abs/2009ApJ...692..324S}
}

@ARTICLE{2009ApJ...705..693S,
       author = {{Shen}, Ken J. and {Idan}, Irit and {Bildsten}, Lars},
        title = "{Helium Core White Dwarfs in Cataclysmic Variables}",
      journal = {\apj},
         year = 2009,
        month = nov,
       volume = {705},
       number = {1},
        pages = {693-703},
          doi = {10.1088/0004-637X/705/1/693},
archivePrefix = {arXiv},
       eprint = {0906.3767},
 primaryClass = {astro-ph.HE},
       adsurl = {https://ui.adsabs.harvard.edu/abs/2009ApJ...705..693S}
}

@ARTICLE{2013ApJ...776...97M,
       author = {{Moore}, Kevin and {Townsley}, Dean M. and {Bildsten}, Lars},
        title = "{The Effects of Curvature and Expansion on Helium Detonations on White Dwarf Surfaces}",
      journal = {\apj},
         year = 2013,
        month = oct,
       volume = {776},
       number = {2},
          eid = {97},
        pages = {97},
          doi = {10.1088/0004-637X/776/2/97},
archivePrefix = {arXiv},
       eprint = {1308.4193},
 primaryClass = {astro-ph.SR},
       adsurl = {https://ui.adsabs.harvard.edu/abs/2013ApJ...776...97M}
}

@ARTICLE{2019ApJ...886...22Z,
       author = {{Zha}, Shuai and {Leung}, Shing-Chi and {Suzuki}, Toshio and {Nomoto}, Ken'ichi},
        title = "{Evolution of ONeMg Core in Super-AGB Stars toward Electron-capture Supernovae: Effects of Updated Electron-capture Rate}",
      journal = {\apj},
         year = 2019,
        month = nov,
       volume = {886},
       number = {1},
          eid = {22},
        pages = {22},
          doi = {10.3847/1538-4357/ab4b4b},
archivePrefix = {arXiv},
       eprint = {1907.04184},
 primaryClass = {astro-ph.HE},
       adsurl = {https://ui.adsabs.harvard.edu/abs/2019ApJ...886...22Z}
}

@ARTICLE{2022Natur.612..223R,
       author = {{Rastinejad}, Jillian C. and others},
        title = "{A kilonova following a long-duration gamma-ray burst at 350 Mpc}",
      journal = {\nat},
         year = 2022,
        month = dec,
       volume = {612},
       number = {7939},
        pages = {223-227},
          doi = {10.1038/s41586-022-05390-w},
archivePrefix = {arXiv},
       eprint = {2204.10864},
 primaryClass = {astro-ph.HE},
       adsurl = {https://ui.adsabs.harvard.edu/abs/2022Natur.612..223R}
}

@article{Levan:2024,
    author = "Levan, Andrew J. and others",
    title = "{Heavy element production in a compact object merger observed by JWST}",
    doi = "10.1038/s41586-023-06759-1",
    journal = "Nature",
    volume = "626",
    pages = "737--741",
    year = "2024"
}

@article{Korobkin:2019uxw,
	author = "Korobkin, Oleg and others",
	title = "{Gamma-rays from kilonova: a potential probe of r-process nucleosynthesis}",
	eprint = "1905.05089",
	archivePrefix = "arXiv",
	primaryClass = "astro-ph.HE",
	reportNumber = "LA-UR-19-24338",
	doi = "10.3847/1538-4357/ab64d8",
	journal = "Astrophys. J.",
	volume = "889",
	pages = "168",
	month = "5",
	year = "2020"
}

@article{Gross:2025yff,
    author = "Gross, Axel and Cupp, Samuel and Mumpower, Matthew R.",
    title = "{Multimessengers from the Radioactive Decay of r-process Nuclei}",
    eprint = "2509.00267",
    archivePrefix = "arXiv",
    primaryClass = "astro-ph.HE",
    reportNumber = "LA-UR-25-28481",
    doi = "10.3847/2041-8213/ae2465",
    journal = "Astrophys. J. Lett.",
    volume = "995",
    number = "1",
    pages = "L28",
    year = "2025"
}

@article{Terada:2022hut,
    author = "Terada, Yukikatsu and Miwa, Yuya and Ohsumi, Hayato and Fujimoto, Shin-ichiro and Katsuda, Satoru and Bamba, Aya and Yamazaki, Ryo",
    title = "{Gamma-ray Diagnostics of r-process Nucleosynthesis in the Remnants of Galactic Binary Neutron-Star Mergers}",
    eprint = "2205.05407",
    archivePrefix = "arXiv",
    primaryClass = "astro-ph.HE",
    doi = "10.3847/1538-4357/ac721f",
    journal = "Astrophys. J.",
    volume = "933",
    pages = "111",
    month = "7",
    year = "2022"
}

@article{Li:2019gamma,
    author = "Li, Li-Xin",
    title = "{Radioactive Gamma-Ray Emissions from Neutron Star Mergers}",
    eprint = "1808.09833",
    archivePrefix = "arXiv",
    primaryClass = "astro-ph.HE",
    doi = "10.3847/1538-4357/aaf961",
    journal = "Astrophys. J.",
    volume = "872",
    pages = "19",
    year = "2019"
}

@article{Lippuner:2017tyn,
	author = "Lippuner, Jonas and Roberts, Luke F.",
	title = "{SkyNet: A modular nuclear reaction network library}",
	eprint = "1706.06198",
	archivePrefix = "arXiv",
	primaryClass = "astro-ph.HE",
	doi = "10.3847/1538-4365/aa94cb",
	journal = "Astrophys. J. Suppl.",
	volume = "233",
	number = "2",
	pages = "18",
	year = "2017"
}

@article{Morozova:2015bla,
    author = "Morozova, Viktoriya and Piro, Anthony L. and Renzo, Mathieu and Ott, Christian D. and Clausen, Drew and Couch, Sean M. and Ellis, Justin and Roberts, Luke F.",
    title = "{Light Curves of Core-Collapse Supernovae with Substantial Mass Loss Using the New Open-Source SuperNova Explosion Code (SNEC)}",
    doi = "10.1088/0004-637X/814/1/63",
    journal = "Astrophys. J.",
    volume = "814",
    pages = "63",
    year = "2015"
}

@article{Wu:2021ibi,
    author = "Wu, Zhenyu and Ricigliano, Giacomo and Kashyap, Rahul and Perego, Albino and Radice, David",
    title = "{Radiation hydrodynamics modelling of kilonovae with SNEC}",
    eprint = "2111.06870",
    archivePrefix = "arXiv",
    primaryClass = "astro-ph.HE",
    doi = "10.1093/mnras/stac399",
    journal = "Mon. Not. Roy. Astron. Soc.",
    volume = "512",
    number = "1",
    pages = "328--347",
    year = "2022"
}

@unpublished{Magistrelli:2025xja,
    author = "Magistrelli, Fabio and Bernuzzi, Sebastiano and Perego, Albino and Jacobi, Maximilian and Fontes, Christopher J.",
    title = "{Impact of in-situ nuclear networks and atomic opacities on neutron star merger ejecta dynamics, nucleosynthesis, and kilonovae}",
    eprint = "2512.11032",
    archivePrefix = "arXiv",
    primaryClass = "astro-ph.HE",
    note = "arXiv:2512.11032",
    year = "2025"
}

@ARTICLE{2020CQGra..37n5015C,
       author = {{Cheong}, Patrick Chi-Kit and {Lin}, Lap-Ming and {Li}, Tjonnie Guang Feng},
        title = "{Gmunu: toward multigrid based Einstein field equations solver for general-relativistic hydrodynamics simulations}",
      journal = {Classical and Quantum Gravity},
         year = 2020,
        month = jul,
       volume = {37},
       number = {14},
        pages = {145015},
          doi = {10.1088/1361-6382/ab8e9c},
archivePrefix = {arXiv},
       eprint = {2001.05723},
 primaryClass = {gr-qc}
}

@ARTICLE{2021MNRAS.508.2279C,
       author = {{Cheong}, Patrick Chi-Kit and {Lam}, Alan Tsz-Lok and {Ng}, Harry Ho-Yin and {Li}, Tjonnie Guang Feng},
        title = "{Gmunu: paralleled, grid-adaptive, general-relativistic magnetohydrodynamics in curvilinear geometries in dynamical space-times}",
      journal = {\mnras},
         year = 2021,
        month = dec,
       volume = {508},
       number = {2},
        pages = {2279--2301},
          doi = {10.1093/mnras/stab2606},
archivePrefix = {arXiv},
       eprint = {2012.07322},
 primaryClass = {astro-ph.IM}
}

@ARTICLE{2018NDS...148....1B,
	author = {{Brown}, D.~A. and {Chadwick}, M.~B. and {Capote}, R. and {Kahler}, A.~C. and {Trkov}, A. and {Herman}, M.~W. and {Sonzogni}, A.~A. and {Danon}, Y. and {Carlson}, A.~D. and {Dunn}, M. and {Smith}, D.~L. and {Hale}, G.~M. and {Arbanas}, G. and {Arcilla}, R. and {Bates}, C.~R. and {Beck}, B. and {Becker}, B. and {Brown}, F. and {Casperson}, R.~J. and {Conlin}, J. and {Cullen}, D.~E. and {Descalle}, M.-A. and {Firestone}, R. and {Gaines}, T. and {Guber}, K.~H. and {Hawari}, A.~I. and {Holmes}, J. and {Johnson}, T.~D. and {Kawano}, T. and {Kiedrowski}, B.~C. and {Koning}, A.~J. and {Kopecky}, S. and {Leal}, L. and {Lestone}, J.~P. and {Lubitz}, C. and {M{\'a}rquez Dami{\'a}n}, J.~I. and {Mattoon}, C.~M. and {McCutchan}, E.~A. and {Mughabghab}, S. and {Navratil}, P. and {Neudecker}, D. and {Nobre}, G.~P.~A. and {Noguere}, G. and {Paris}, M. and {Pigni}, M.~T. and {Plompen}, A.~J. and {Pritychenko}, B. and {Pronyaev}, V.~G. and {Roubtsov}, D. and {Rochman}, D. and {Romano}, P. and {Schillebeeckx}, P. and {Simakov}, S. and {Sin}, M. and {Sirakov}, I. and {Sleaford}, B. and {Sobes}, V. and {Soukhovitskii}, E.~S. and {Stetcu}, I. and {Talou}, P. and {Thompson}, I. and {van der Marck}, S. and {Welser-Sherrill}, L. and {Wiarda}, D. and {White}, M. and {Wormald}, J.~L. and {Wright}, R.~Q. and {Zerkle}, M. and {{\v{Z}}erovnik}, G. and {Zhu}, Y.},
	title = "{ENDF/B-VIII.0: The 8$^{th}$ Major Release of the Nuclear Reaction Data Library with CIELO-project Cross Sections, New Standards and Thermal Scattering Data}",
	journal = {Nuclear Data Sheets},
	year = 2018,
	month = feb,
	volume = {148},
	pages = {1-142},
	doi = {10.1016/j.nds.2018.02.001},
	adsurl = {https://ui.adsabs.harvard.edu/abs/2018NDS...148....1B}
}

@ARTICLE{2010ApJS..189..240C,
       author = {{Cyburt}, R.~H. and others},
        title = "{The JINA REACLIB Database: Its Recent Updates and Impact on Type-I X-Ray Bursts}",
      journal = {\apjs},
         year = 2010,
        month = aug,
       volume = {189},
       number = {1},
        pages = {240-252},
          doi = {10.1088/0067-0049/189/1/240},
archivePrefix = {arXiv},
       eprint = {1005.3131}
}

@article{Moller:2015fba,
    author = "M{\"o}ller, Peter and Sierk, Arnold J. and Ichikawa, Takatoshi and Sagawa, Hiroyuki",
    title = "{Nuclear ground-state masses and deformations: FRDM(2012)}",
    doi = "10.1016/j.adt.2015.10.002",
    journal = "At. Data Nucl. Data Tables",
    volume = "109-110",
    pages = "1--204",
    year = "2016"
}

@ARTICLE{1991NuPhA.535..331L,
       author = {{Lattimer}, James M. and {Swesty}, F. Douglas},
        title = "{A generalized equation of state for hot, dense matter}",
      journal = {Nuclear Physics A},
         year = 1991,
        month = dec,
       volume = {535},
       number = {2},
        pages = {331-376},
          doi = {10.1016/0375-9474(91)90452-C}
}

@article{Tomsick:2023aue,
    author = "Tomsick, John A. and others",
    title = "{The Compton Spectrometer and Imager}",
    eprint = "2308.12362",
    archivePrefix = "arXiv",
    primaryClass = "astro-ph.HE",
    doi = "10.22323/1.444.0745",
    journal = "PoS",
    volume = "ICRC2023",
    pages = "745",
    year = "2023"
}

@article{e-ASTROGAM:2017pxr,
    author = "Tavani, M. and others",
    editor = "De Angelis, A. and Tatischeff, V. and Grenier, I. A. and McEnery, J. and Mallamaci, M.",
    collaboration = "e-ASTROGAM",
    title = "{Science with e-ASTROGAM: A space mission for MeV{\textendash}GeV gamma-ray astrophysics}",
    eprint = "1711.01265",
    archivePrefix = "arXiv",
    primaryClass = "astro-ph.HE",
    doi = "10.1016/j.jheap.2018.07.001",
    journal = "JHEAp",
    volume = "19",
    pages = "1--106",
    year = "2018"
}

@unpublished{Shutt:2025xvc,
    author = "Shutt, Tom and others",
    title = "{The GammaTPC Gamma-Ray Telescope Concept}",
    eprint = "2502.14841",
    archivePrefix = "arXiv",
    primaryClass = "astro-ph.IM",
    reportNumber = "FERMILAB-PUB-25-0108-PPD",
    note = "arXiv:2502.14841",
    month = "2",
    year = "2025"
}

@unpublished{Combi:2025yvs,
    author = "Combi, Luciano and Siegel, Daniel M. and Metzger, Brian D.",
    title = "{Jet-driven explosion of an accretion-induced white-dwarf collapse via a magnetorotational dynamo}",
    eprint = "2509.19799",
    archivePrefix = "arXiv",
    primaryClass = "astro-ph.HE",
    note = "arXiv:2509.19799",
    month = "9",
    year = "2025"
}

@article{Chen:2021tob,
	author = "Chen, Meng-Hua and Li, Li-Xin and Lin, Da-Bin and Liang, En-Wei",
	title = "{Gamma-Ray Emission Produced by r-process Elements from Neutron Star Mergers}",
	eprint = "2107.02982",
	archivePrefix = "arXiv",
	primaryClass = "astro-ph.HE",
	doi = "10.3847/1538-4357/ac1267",
	journal = "Astrophys. J.",
	volume = "919",
	number = "1",
	pages = "59",
	year = "2021"
}

@ARTICLE{1975JPCRD...4..471H,
       author = {{Hubbell}, J.~H. and {Veigele}, Wm. J. and {Briggs}, E.~A. and {Brown}, R.~T. and {Cromer}, D.~T. and {Howerton}, R.~J.},
        title = "{Atomic form factors, incoherent scattering functions, and photon scattering cross sections}",
      journal = {Journal of Physical and Chemical Reference Data},
         year = 1975,
        month = jul,
       volume = {4},
       number = {3},
        pages = {471-538},
          doi = {10.1063/1.555523},
       adsurl = {https://ui.adsabs.harvard.edu/abs/1975JPCRD...4..471H}
}

@ARTICLE{2007ApJ...669..585D,
       author = {{Dessart}, L. and {Burrows}, A. and {Livne}, E. and {Ott}, C.~D.},
        title = "{Magnetically Driven Explosions of Rapidly Rotating White Dwarfs Following Accretion-Induced Collapse}",
      journal = {\apj},
         year = 2007,
        month = nov,
       volume = {669},
       number = {1},
        pages = {585-599},
          doi = {10.1086/521701},
archivePrefix = {arXiv},
       eprint = {0705.3678},
 primaryClass = {astro-ph},
       adsurl = {https://ui.adsabs.harvard.edu/abs/2007ApJ...669..585D}
}

@article{Batziou:2024ory,
	author = "Batziou, Eirini and Glas, Robert and Janka, H. -Thomas and Ehring, Jakob and Abdikamalov, Ernazar and Just, Oliver",
	title = "{Nucleosynthesis Conditions in Outflows of White Dwarfs Collapsing to Neutron Stars}",
	eprint = "2412.02756",
	archivePrefix = "arXiv",
	primaryClass = "astro-ph.HE",
	doi = "10.3847/1538-4357/adc300",
	journal = "Astrophys. J.",
	volume = "984",
	number = "2",
	pages = "197",
	year = "2025"
}

@ARTICLE{1985A&A...150L..21S,
	author = {{Saio}, H. and {Nomoto}, K.},
	title = "{Evolution of a merging pair of C + O white dwarfs to form a single neutron star}",
	journal = {\aap},
	year = 1985,
	month = sep,
	volume = {150},
	number = {1},
	pages = {L21-L23},
	adsurl = {https://ui.adsabs.harvard.edu/abs/1985A&A...150L..21S}
}

@article{Wang:2020pzc,
	author = "Wang, Bo and Liu, Dongdong",
	title = "{The formation of neutron star systems through accretion-induced collapse in white-dwarf binaries}",
	eprint = "2005.01880",
	archivePrefix = "arXiv",
	primaryClass = "astro-ph.SR",
	doi = "10.1088/1674-4527/20/9/135",
	journal = "Res. Astron. Astrophys.",
	volume = "20",
	number = "9",
	pages = "135",
	year = "2020"
}

@ARTICLE{2006MNRAS.368L...1L,
	author = {{Levan}, Andrew J. and {Wynn}, Graham A. and {Chapman}, Robert and {Davies}, Melvyn B. and {King}, Andrew R. and {Priddey}, Robert S. and {Tanvir}, Nial R.},
	title = "{Short gamma-ray bursts in old populations: magnetars from white dwarf-white dwarf mergers}",
	journal = {\mnras},
	year = 2006,
	month = may,
	volume = {368},
	number = {1},
	pages = {L1-L5},
	doi = {10.1111/j.1745-3933.2006.00144.x},
	archivePrefix = {arXiv},
	eprint = {astro-ph/0601332},
	primaryClass = {astro-ph},
	adsurl = {https://ui.adsabs.harvard.edu/abs/2006MNRAS.368L...1L}
}

@article{Kremer:2023utr,
    author = "Kremer, Kyle and Fuller, Jim and Piro, Anthony L. and Ransom, Scott M.",
    title = "{Connecting the young pulsars in Milky Way globular clusters with white dwarf mergers and the M81 fast radio burst}",
    eprint = "2305.11933",
    archivePrefix = "arXiv",
    primaryClass = "astro-ph.HE",
    doi = "10.1093/mnrasl/slad088",
    journal = "Mon. Not. Roy. Astron. Soc.",
    volume = "525",
    number = "1",
    pages = "L22--L27",
    year = "2023"
}

@article{Freire:2013xma,
    author = "Freire, Paulo C. C. and Tauris, Thomas M.",
    title = "{Direct formation of millisecond pulsars from rotationally delayed accretion-induced collapse of massive white dwarfs}",
    eprint = "1311.3478",
    archivePrefix = "arXiv",
    primaryClass = "astro-ph.SR",
    doi = "10.1093/mnrasl/slt164",
    journal = "Mon. Not. Roy. Astron. Soc.",
    volume = "438",
    pages = "86",
    year = "2014"
}

@ARTICLE{1984JApA....5..209V,
       author = {{van den Heuvel}, E.~P.~J.},
        title = "{Models for the formation of binary and millisecond radio pulsars}",
      journal = {Journal of Astrophysics and Astronomy},
         year = 1984,
        month = sep,
       volume = {5},
       number = {3},
        pages = {209-233},
          doi = {10.1007/BF02714540},
       adsurl = {https://ui.adsabs.harvard.edu/abs/1984JApA....5..209V}
}

@article{Hotokezaka:2015cma,
	author = "Hotokezaka, Kenta and Wanajo, Shinya and Tanaka, Masaomi and Bamba, Aya and Terada, Yukikatsu and Piran, Tsvi",
	title = "{Radioactive decay products in neutron star merger ejecta: heating efficiency and {\ensuremath{\gamma}}-ray emission}",
	eprint = "1511.05580",
	archivePrefix = "arXiv",
	primaryClass = "astro-ph.HE",
	doi = "10.1093/mnras/stw404",
	journal = "Mon. Not. Roy. Astron. Soc.",
	volume = "459",
	number = "1",
	pages = "35--43",
	year = "2016"
}

@unpublished{Jacobi:2025eak,
    author = "Jacobi, Maximilian and Magistrelli, Fabio and Loffredo, Eleonora and Ricigliano, Giacomo and Chiesa, Leonardo and Bernuzzi, Sebastiano and Perego, Albino and Arcones, Almudena",
    title = "{$^{56}$Ni production in long-lived binary neutron star merger remnants}",
    eprint = "2503.17445",
    archivePrefix = "arXiv",
    primaryClass = "astro-ph.HE",
    note = "arXiv:2503.17445",
    month = "3",
    year = "2025"
}

@unpublished{Pitik:2026bjm,
    author = "Pitik, Tetyana and Radice, David and Kasen, Daniel and Magistrelli, Fabio and Cheong, Patrick Chi-Kit and Bernuzzi, Sebastiano",
    title = "{Collapse of Magnetized White Dwarfs as site of Heavy Element Formation and Kilonova Signal}",
    eprint = "2602.21291",
    archivePrefix = "arXiv",
    primaryClass = "astro-ph.HE",
    note = "arXiv:2602.21291",
    month = "2",
    year = "2026"
}

@inproceedings{GRAMS:2025ljc,
    author = "Zeng, J. and others",
    collaboration = "GRAMS",
    title = "{Gamma-Ray and AntiMatter Survey (GRAMS) experiment}",
    booktitle = "{19th International Conference on Topics in Astroparticle and Underground Physics}",
    eprint = "2512.14913",
    archivePrefix = "arXiv",
    primaryClass = "hep-ex",
    month = "12",
    year = "2025"
}

@article{Caputo:2022xpx,
    author = "Caputo, Regina and others",
    title = "{All-sky Medium Energy Gamma-ray Observatory eXplorer mission concept}",
    eprint = "2208.04990",
    archivePrefix = "arXiv",
    primaryClass = "astro-ph.IM",
    doi = "10.1117/1.JATIS.8.4.044003",
    journal = "J. Astron. Telesc. Instrum. Syst.",
    volume = "8",
    number = "4",
    pages = "044003",
    year = "2022"
}

@ARTICLE{2009MNRAS.396.1659M,
       author = {{Metzger}, B.~D. and {Piro}, A.~L. and {Quataert}, E.},
        title = "{Nickel-rich outflows from accretion discs formed by the accretion-induced collapse of white dwarfs}",
      journal = {\mnras},
         year = 2009,
        month = jul,
       volume = {396},
       number = {3},
        pages = {1659-1664},
          doi = {10.1111/j.1365-2966.2009.14909.x},
archivePrefix = {arXiv},
       eprint = {0812.3656},
 primaryClass = {astro-ph},
       adsurl = {https://ui.adsabs.harvard.edu/abs/2009MNRAS.396.1659M}
}

\end{document}